\documentclass[  superscriptaddress, amsmath,amssymb, aps,twocolumn,longbibliography]{revtex4-1} 
\usepackage{graphicx}
\usepackage{dcolumn}
\usepackage[T1]{fontenc}
\usepackage{newunicodechar}

\newunicodechar{−}{\ensuremath{-}}
\newunicodechar{≈}{\ensuremath{\approx}}
\usepackage{amsthm}
\usepackage{bm}

\pdfoutput=1
\usepackage{graphicx,graphics,epsfig,subfigure,times,bm,bbm,amssymb,amsmath,amsthm,mathrsfs,MnSymbol}
\usepackage{gensymb}
\usepackage{amsfonts}
\usepackage[matrix,frame,arrow]{xypic}
\usepackage[pdfstartview=FitH]{hyperref}
\hypersetup{
    colorlinks=true,       
    linkcolor=red,          
    citecolor=magenta,        
    filecolor=magenta,      
    urlcolor=cyan,           
    runcolor=cyan}
\usepackage{epstopdf}
\usepackage[pdftex]{color}
\usepackage{braket}
\usepackage{enumerate}
\usepackage[normalem]{ulem}
\usepackage[usenames,dvipsnames]{xcolor}
\usepackage{multirow}
\usepackage{mathtools}
\usepackage{changes}

\definecolor{orange}{rgb}{1,0.5,0}

\begin{document}

\title{Evidence for a Dirac Spin Liquid in the Kagome Heisenberg Antiferromagnet from High-Precision Neural-Network Quantum States}

\author{Zheng-Hang Sun}
\affiliation{Theoretical Physics \uppercase\expandafter{\romannumeral3}, Center for Electronic Correlations and Magnetism, Institute of Physics, University of Augsburg, 86135 Augsburg, Germany}

\author{Christopher Roth}
\affiliation{Stewart Blusson Quantum Matter Institute, University of British Columbia, Vancouver, BC V6T 1Z4, Canada}
\affiliation{Center for Computational Quantum Physics, Flatiron Institute, 162 5th Avenue, New York, NY 10010, USA}

\author{Ao Chen}
\email{chenao.phys@gmail.com}
\affiliation{Division of Chemistry and Chemical Engineering, California Institute of Technology, Pasadena, California 91125, USA}

\author{Markus Heyl}
\email{markus.heyl@uni-a.de}
\affiliation{Theoretical Physics \uppercase\expandafter{\romannumeral3}, Center for Electronic Correlations and Magnetism, Institute of Physics, University of Augsburg, 86135 Augsburg, Germany}
\affiliation{Centre for Advanced Analytics and Predictive Sciences (CAAPS),
University of Augsburg, Universitätsstr. 12a, 86159 Augsburg, Germany}

\begin{abstract}
\noindent The spin-1/2 kagome Heisenberg antiferromagnet is a central model in the study of quantum spin liquids, yet its ground-state properties remain unresolved. 
We perform large-scale variational Monte-Carlo calculations using neural-network quantum states (NQSs) with kagome-geometry-aware message-passing convolution on two-dimensional tori containing up to 300 spins, and achieve exceptionally low variational energies for the kagome model. The spin and dimer structure factors show no evidence of magnetic or valence-bond long-range order. Moreover, the dominant symmetry-resolved spin and dimer correlations exhibit a common scaling behavior consistent with the emergent $SO(6)$ monopole multiplet of the Dirac spin liquid. Finally, the fidelity of the optimized NQSs with Gutzwiller-projected mean-field states increasingly favors the Dirac state over the competing $Z_{2}$ candidates as the system size grows. Together, these complementary results provide evidence for a $U(1)$ Dirac spin-liquid ground state in the kagome Heisenberg antiferromagnet. 
\end{abstract}
\pacs{Valid PACS appear here}
\maketitle


\emph{Introduction.}--Quantum spin liquids (QSLs) represent a fundamental class of quantum phases beyond the Landau symmetry-breaking paradigm, exhibiting fractionalized excitations, long-ranged quantum entanglement, and emergent gauge fields~\cite{Balents:2010ve,Savary_2017,RevModPhys.89.025003,Broholm2020QuantumSpinLiquids}. Understanding the nature of such phases of matter has become one of the central questions in condensed-matter physics. The long-standing experimental search for QSLs has placed kagome antiferromagnets at the forefront of two-dimensional quantum magnetism, with prominent candidate materials including herbertsmithite~\cite{RevModPhys.88.041002,10.1021/ja053891p,PhysRevLett.98.107204,Han:2012wk}, Zn-barlowite~\cite{Feng_2017,Wang:2021up,Breidenbach:2025to}, and Y-based kagome antiferromagnets~\cite{CHEN2020167066,Jeon:2024wh,Zeng:2024ui}. The spin-1/2 kagome Heisenberg model with nearest-neighbour interactions has become the minimal canonical model for kagome antiferromagnets, yet the nature of its ground state remains elusive. Early density-matrix renormalization-group (DMRG) calculations provided evidence for a $Z_{2}$ gapped QSL~\cite{doi:10.1126/science.1201080,PhysRevLett.109.067201,Jiang:2012te}. However, subsequent calculations based on infinite projected entangled simplex states found scaling behaviour consistent with a gapless state~\cite{PhysRevLett.118.137202}, and flux-threaded infinite-cylinder DMRG revealed Dirac-cone signatures~\cite{PhysRevX.7.031020}. Alongside tensor-network simulations, variational Monte Carlo (VMC) studies based on Gutzwiller-projected fermionic mean-field ansätze initially identified the $\pi$-flux U(1) Dirac QSL as a leading variational candidate~\cite{PhysRevLett.98.117205,PhysRevB.84.020407,PhysRevB.87.060405,PhysRevB.89.020407,PhysRevB.104.144406}, although further optimizations have also revealed closely competing $Z_{2}$ states~\cite{PhysRevB.102.195106}.


 
Despite the aforementioned numerical progress, both DMRG and VMC face intrinsic limitations. DMRG is largely restricted to quasi-one-dimensional cylinders, with a cost that grows rapidly with circumference, whereas VMC is constrained by the bias and limited expressivity of the underlying mean-field ansatz. Neural-network quantum states (NQSs) provide a variational framework that can be directly applied to large two-dimensional lattices with periodic boundary conditions, thereby allowing studies on torus rather than quasi-one-dimensional cylinders. Moreover, NQSs can represent complex many-body correlations and sign structures without being constrained to a specific parton mean-field manifold~\cite{doi:10.1126/science.aag2302,PhysRevX.7.021021,Gao:2017uk}. 
Recent NQS-based studies have achieved 
ground-state energy estimates competitive with the best available results
for $J_{1}-J_{2}$ square-lattice and triangular Heisenberg models~\cite{PhysRevX.11.031034,PhysRevB.108.054410,Chen:2024tb,7xwp-25y9,viteritti2026approachingthermodynamiclimitneuralnetwork}, as well as two-dimensional Fermi-Hubbard models~\cite{doi:10.1073/pnas.2535288123,roth2025superconductivitytwodimensionalhubbardmodel,Gu:2026vw,rende2026superconductivitytthubbardmodel}. 

The group convolutional neural-network (G-CNN) architecture of NQS has been applied to the kagome Heisenberg antiferromagnet, suggesting an evidence of spinon pair-density-wave (SPDW) ground state~\cite{PhysRevX.15.011047}. However, a subsequent analysis reveals that the unusually low variational energies reported in Ref.~\cite{PhysRevX.15.011047} are artifacts induced by nonergodic Monte Carlo sampling~\cite{kamal2026commentspin12kagomeheisenberg}, and the signatures of SPDW require a further demonstration. 
These deficiencies leave the NQS study of the kagome Heisenberg ground state unresolved and demand a rigorous reassessment based on reliable sampling.

\begin{figure*}[]
	\centering
	\includegraphics[width=1\linewidth]{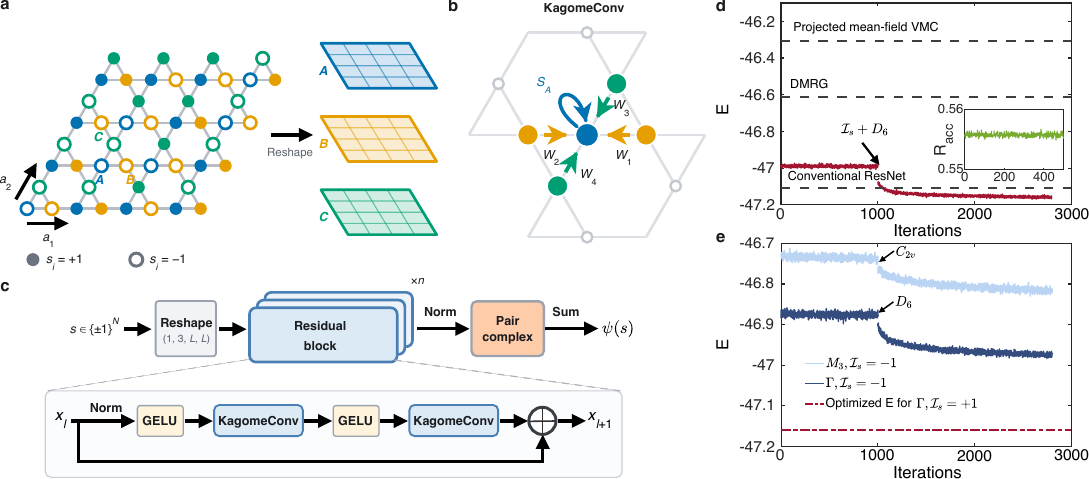}\\
	\caption{\textbf{The kagome-geometry-aware neural quantum state and variational energy optimization for $L=6$.} \textbf{a}, Input encoding. A spin configuration $\mathbf{s}\in \{\pm 1 \}^{N}$  on the kagome lattice, with $N = 3L^{2}$ sites (shown for $L=4$ as an example), is reshaped into a tensor of shape $(1,3,L,L)$. The three sublattice sites $A$, $B$ and $C$ within each unit cell form three $L\times L$ feature maps indexed along the primitive vectors $\mathbf{a}_{1}$ and $\mathbf{a}_{2}$ of the underlying triangular Bravais lattice. Periodic boundary conditions are implemented through cyclic shifts of the Bravais-lattice indices.  \textbf{b}, Kagome message-passing convolution (KagomeConv). Each site receives messages from its four nearest neighbors and from itself, illustrated for an $A$-sublattice site as an example. Each directed nearest-neighbor bond class carries an independent kernel $W_e \in \mathbb{R}^{c\times c}$, with $c$ being the number of feature channels. Here we plot $W_{1,2,3,4}$ for an $A$-site as an example, and there are 12 bond classes in total for a unit cell. Each sublattice $\alpha$ has its own on-site kernel $S_{\alpha}$ and bias. \textbf{c}, Network architecture of  KagomeResNet. The input is processed by $n$ residual blocks, each consisting of normalization (Norm), GELU activations, and two layers of KagomeConv connected by a residual shortcut. After a final normalization, the pair-complex activation combines real channels into complex outputs, which are summed to yield the wave-function amplitude $\psi(\mathbf{s})$.  \textbf{d}, Variational-energy optimization for $L=6$ in the $\Gamma$-point sector. After optimization with translational symmetry, the spin-inversion symmetry $\mathcal{I}_{s}=+1$ and the $D_{6}$ point-group symmetry are imposed at the arrow-marked iteration. The inset shows the Monte Carlo acceptance rate $R_{\mathrm{acc}}$ in the final 500 steps of the symmetry-projected stage. For comparison, we highlight the raw energies obtained by projected VMC based on $U(1)$ Dirac QSL $E\simeq -46.307$~\cite{PhysRevB.87.060405,PhysRevB.89.020407}, DMRG $E\simeq -46.613$~\cite{PhysRevLett.109.067201,PhysRevB.87.060405}, and conventional ResNet $E\simeq -47.105$ (see the Supplementary Material) by dashed lines.  \textbf{e}, Variational-energy optimization for $L=6$ in the
odd spin-inversion sector $\mathcal{I}_{s}=-1$ at the $M_{3}$ and
$\Gamma$ points. The corresponding little-group symmetries,
$C_{2v}$ and $D_{6}$, are imposed at the iteration marked by the
arrows. The dashed red line marks the optimized variational energy $E\simeq -47.160$ in the
$\Gamma$-point sector with $\mathcal{I}_{s}=+1$. }\label{fig1}
\end{figure*}

In this work, we study ground-state properties of spin-1/2 kagome Heisenberg antiferromagnet using deep neural quantum states whose architecture is tailored to the kagome geometry. Our variational ansatz extends the residual convolutional network architecture developed in Ref.~\cite{PhysRevB.108.054410,Chen:2024tb,7xwp-25y9}, i.e., the ResNet, to the kagome model by replacing 
conventional grid-based kernels with message-passing convolutions that propagate features along the nearest-neighbour bonds while retaining an explicit representation of the three sublattices in each unit cell (see Fig.~\ref{fig1}\textbf{a}-\textbf{c}). We benchmark that the ResNet with geometry-aware message-passing convolutions (KagomeResNet) outperforms its standard-convolution counterparts. To the best of our knowledge, the resulting variational energies for large system sizes on a torus achieve the lowest raw energies, apart from the disputed G-CNN results of Ref.~\cite{PhysRevX.15.011047}.
More importantly, based on the analysis of static structure factors, correlation functions, and projections of the optimized NQS onto manifolds of Gutzwiller-projected mean-field states, we provide consistent evidence for a Dirac QSL ground state in the spin-1/2 kagome Heisenberg antiferromagnet.

\emph{Model and variational wave function.}--The Hamiltonian of kagome Heisenberg antiferromagnet studied in this work is 
\begin{eqnarray}
\mathcal{H} = \sum_{\langle i,j \rangle} \mathbf{S}_{i} \cdot \mathbf{S}_{j}
\label{eq_h}
\end{eqnarray}
with $\mathbf{S}_{i} = (S_{i}^{x},S_{i}^{y},S_{i}^{z})$ being the spin-1/2 operators, and $\langle i,j \rangle$ indicating pairs of nearest-neighbor sites on a kagome lattice shown in Fig.~\ref{fig1}\textbf{a}. Here, we consider the periodic boundary condition of the model, i.e., the torus, and the system sizes $L=6$, $8$ and $10$, with number of sites $N = 108$, $192$ and $300$, respectively ($N=L\times L\times 3$). 

In the NQS approach of training the ground state of $\mathcal{H}$, a neural network provides a compact parametrization of the many-body wave function. For a spin-1/2 system with $N$ sites, the Hilbert space is spanned by the $z$-basis spin configurations $|s\rangle = |s_{1},...,s_{N}\rangle$ with $s_{i} = \pm 1$. The NQS parameterized by $\boldsymbol{\theta}$ can map each spin configuration $s$ to a complex wave-function amplitude $\psi_{\boldsymbol{\theta}}(s)$. To provide a phase prior for the ground state of Hamiltonian~(\ref{eq_h}),, we incorporate a fixed reference phase corresponding to a coplanar $q=0$ three-sublattice pattern $\phi_{q=0}(s)$ into the wave-function parametrization, thereby defining the variational state $|\Psi_{\boldsymbol{\theta}}\rangle = \sum_{s} \psi_{\boldsymbol{\theta}}(s)\phi_{q=0}(s) |s\rangle$. 
Within variational Monte Carlo, the parameters $\boldsymbol{\theta}$ are optimized using MinSR~\cite{Chen:2024tb}, to minimize the variational energy $E_{\boldsymbol{\theta}} = \langle \Psi_{\boldsymbol{\theta}} | \mathcal{H} | \Psi_{\boldsymbol{\theta}} \rangle /\langle \Psi_{\boldsymbol{\theta}} | \Psi_{\boldsymbol{\theta}} \rangle $. We emphasize that the complex-valued network $\psi_{\boldsymbol{\theta}}(s)$ can learn additional phase corrections beyond the reference factor $\phi_{q=0}(s)$. 
More details of the optimization are provided in the Supplemental Material.

The architecture of our NQS, the KagomeResNet, is plotted in Fig.~\ref{fig1}\textbf{a}-\textbf{c}. First, the spin configuration $s_{i}$ is reshaped into three $L\times L$ matrices associated with the $A$, $B$, and $C$ sublattices of the kagome lattice. Then, as a key operation for our NQS, we perform message passing directly along the nearest-neighbor kagome connectivity. 
Specifically, in Fig.~\ref{fig1}\textbf{b}, we show the operation for an $A$-sublattice site as an example, i.e., 
\begin{eqnarray}
\textbf{y}_{A,i} = S_{A} \textbf{x}_{A,i} + W_{1}\textbf{x}_{B,j_{1}} + W_{2}\textbf{x}_{B,j_{2}} +  W_{3}\textbf{x}_{C,j_{3}} + W_{4}\textbf{x}_{C,j_{4}},
\label{eq_mp}
\end{eqnarray}
where $\textbf{x}_{\alpha,i}$ and $\textbf{y}_{\alpha,i}$ are the input and output feature vectors at site $i$ of sublattice $\alpha$, respectively. The matrices $S_{A}$ and $W_{1,...,4}$ represent trainable transformations for the central site and its four directed nearest-neighbor bonds. Similar operations are performed for the $B$ and $C$ sublattices, and thus there are 12 
nearest-neighbor message-passing channels in a complete kagome convolutional (KagomeConv) layer.  The full structure of KagomeResNet is shown in Fig.~\ref{fig1}\textbf{c}. The $l$-th residual block acts on $\textbf{x}_{l} \in \mathbb{R}^{3 \times L \times L \times c}$, where $c$ is the channel dimension, and maps it to $\textbf{x}_{l+1}$ via two KagomeConv layers with GELU activations and a residual connection. We provide more details for the KagomeResNet and conventional ResNet in the Supplementary Material. The network contains $n$ residual blocks, and the output is finally converted by normalization and a pair-complex activation into the wave-function amplitude $\psi_{\boldsymbol{\theta}}(s) $. 

\emph{Benchmarks.}--We now benchmark the performance of KagomeResNet for training the ground state of the Hamiltonian (\ref{eq_h}) with $L=6$ in the $\Gamma$-point sector. We first perform $10^{4}$ optimization steps imposing only translational symmetry, and subsequently for 1800 steps including the even spin-inversion sector $\mathcal{I}_{s} = +1$ and $D_{6}$  point-group symmetry.  In Fig.~\ref{fig1}\textbf{d}, for the KagomeResNet with $c=n=32$, we show the final 1000 iterations of the translationally symmetric optimization, followed by the full 1800-step optimization after the additional symmetry projections are imposed, and demonstrate that the converged value of the ground-state energy is $E = -47.1597(6)$, which is highly competitive with the previously reported raw variational energies~\cite{PhysRevLett.109.067201,PhysRevB.87.060405,PhysRevB.89.020407}, and also lower than the conventional ResNet comparable number of variational parameters (see Supplemental Material).

Recent work using a G-CNN ansatz for the kagome model (\ref{eq_h}) suggests that non-ergodic Monte Carlo sampling can induce low energy during the optimization, and reports $E\simeq -45.6$ for G-CNN with ergodic sampling~\cite{kamal2026commentspin12kagomeheisenberg}. In the inset of Fig.~\ref{fig1}\textbf{d}, we show the acceptance rate $R_{\text{acc}}\simeq 0.555$ over the final 500 iterations with full symmeries, demonstrating that our Monte Carlo sampling is ergodic. 
The acceptance rate is monitored throughout all subsequent optimizations in this work to ensure that the Monte Carlo sampling remains ergodic. 

We then benchmark the low-lying energies in the odd spin-inversion sector $\mathcal{I}_{s} = -1$. 
For $L=4$, the exact diagonalization study shows that the lowest-energy with $\mathcal{I}_{s} = -1$ lies in the $M$-point sector, while the lowest $\Gamma$-point energy is nearly degenerate, only $\sim 6.8\times 10^{-3}$ higher in energy~\cite{PhysRevB.100.155142}. In Fig.~\ref{fig1}\textbf{e}, we plot the final 1000 iterations of the translationally symmetric optimization with $\mathcal{I}_{s} = -1$ for both $M$ and $\Gamma$-point sectors,  followed by \(1800\) additional iterations imposing their respective little-group symmetries $D_{6}$ and $C_{2v}$. Our results show that, different from the $L=4$ case, the lowest-energy state within the odd spin-inversion sector $\mathcal{I}_{s} = -1$ occurs at the $\Gamma$ point for $L=6$. More importantly, upon comparing both spin-inversion sectors, we find the overall ground state at the $\Gamma$ point with $\mathcal{I}_{s} = +1$. This restores the same ground-state quantum numbers as found by exact diagonalization for $L=4$, in stark contrast to Ref.~\cite{PhysRevX.15.011047}, which reported a change of the ground-state sector to the $M$ point with $\mathcal{I}_{s} = -1$ for $L=6$. We therefore find no evidence for the size-driven change of ground-state quantum numbers that underlies the SPDW interpretation of Ref.~\cite{PhysRevX.15.011047}. 

In Table~\ref{tab:energy}, we systematically represent the energy density $e=E/N$ and the energy density variance $\sigma^{2}$ obtained from KagomeResNet with $n=c=32$ and $n=c=40$ for the kagome model (\ref{eq_h}) with $L=6$, $8$ and $10$ at the $\Gamma$ point with $\mathcal{I}_{s} = +1$.  For the $L=8$ torus, previous references of the raw energy density include $e = -0.4286749(4)$ and $-0.428783(1)$ for the projected mean-field VMC based on $U(1)$ Dirac~\cite{PhysRevB.87.060405,PhysRevB.89.020407} and $Z_{2}$~\cite{PhysRevB.102.195106} QSLs, respectively, and $e\sim -0.4349$ for the symmetric projected entangled-pair states calculations~\cite{SciPostPhys.7.1.006}. Our variational energies are substantially lower than the aforementioned benchmarks, improving the best known estimate for the energy density $e$ to the third decimal place.


\begin{table}[t]
\caption{
Variational energy density $e$ and energy-variance density $\sigma^2$
for KagomeResNet with different system and network sizes.
}
\label{tab:energy}
\begin{ruledtabular}
\begin{tabular}{c cc cc}
& \multicolumn{2}{c}{$c=n=32$}
& \multicolumn{2}{c}{$c=n=40$} \\
$L$ & $e$ & $\sigma^2$ & $e$ & $\sigma^2$ \\
\hline
6  & $-0.436664(6)$ & $0.00350(4)$
   & $-0.436760(5)$ & $0.00312(3)$ \\
8  & $-0.435835(4)$ & $0.00519(5)$
   & $-0.435949(5)$ & $0.00444(4)$ \\
10 & $-0.435239(4)$ & $0.00506(6)$
   & $-0.435472(5)$ & $0.00434(4)$ \\
\end{tabular}
\end{ruledtabular}
\end{table}

\emph{Ground-state correlation functions.}-- To study the ground-state properties of the Hamiltonian (\ref{eq_h}), for the trained NQS with $n=c=40$, we first calculate the spin structure factors \begin{eqnarray}
S^{zz}_{f} (\mathbf{q}) = \frac{1}{N}  \sum_{m,n} e^{i \mathbf{q} \cdot (\mathbf{r}_{m} - \mathbf{r}_{n})}  \langle S_{m}^{z}S_{n}^{z} \rangle,  
\label{spin}
\end{eqnarray}
with $\textbf{q} = (q_{x},q_{y})$ the wave vector in the Cartesian reciprocal-space coordinates, and the dimer structure factors
\begin{eqnarray}
S^{D}_{f} (\mathbf{q}) = \frac{1}{N} \sum_{m,n,k,l} e^{i \mathbf{q} \cdot (\mathbf{r}_{mk} - \mathbf{r}_{nl})} D_{mk,nl}
\label{dimer}
\end{eqnarray}
with the sum running over pairs of nearest-neighbor bonds, $\mathbf r_{mk}=(\mathbf r_m+\mathbf r_k)/2$ being the midpoint of bond $(m,k)$, and $D_{mk,nl} = \langle B_{mk} B_{nl}\rangle - \langle B_{mk}\rangle \langle B_{nl}\rangle$, where $B_{mk}=\mathbf S_m\cdot\mathbf S_k$ is the nearest-neighbor bond-energy operator.

\begin{figure*}[]
	\centering
	\includegraphics[width=1\linewidth]{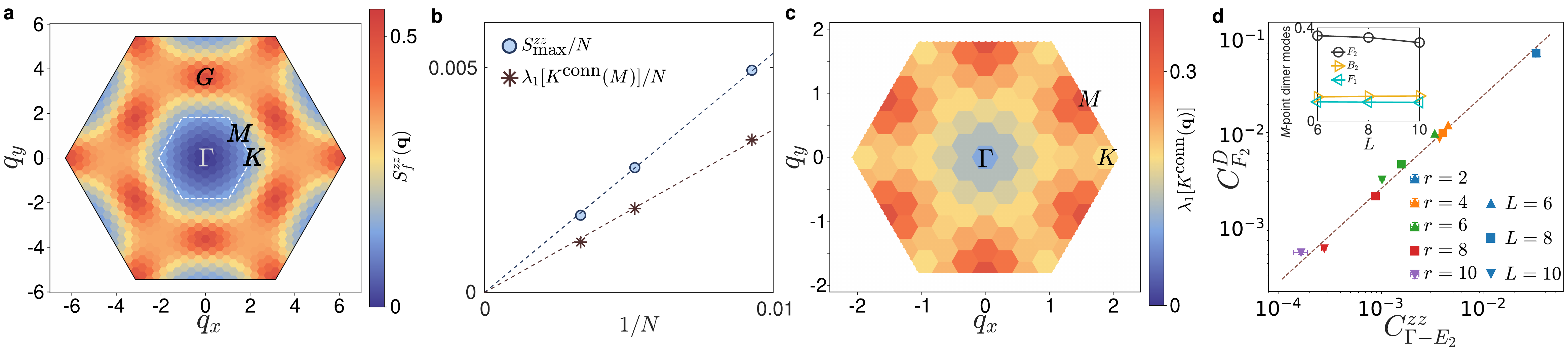}\\
	\caption{\textbf{Structure factors and symmetry-resolved correlations.} \textbf{a},  Spin structure factor $S_f^{zz}(\mathbf q)$ for $L=10$
over an extended Brillouin zone. The dashed hexagon marks the first
Brillouin zone. The maxima lie at the reciprocal-lattice images
$\mathbf G$ of $\Gamma$ probe the nonuniform $\Gamma$-$E_2$ spin
channel. \textbf{b}, Finite-size scaling of the normalized maxima
$S_{\max}^{zz}/N$ and 
$\lambda_1[K^{\rm conn}(M)]/N$. Dashed lines are linear fits through the origin. \textbf{c}, Largest eigenvalue
$\lambda_1[K^{\rm conn}(\mathbf q)]$ of the connected bond-correlation
matrix for $L=10$. The strongest dimer fluctuations occur at the
three symmetry-related $M$ points. \textbf{d}, Correlation between the symmetry-resolved
$M$-$F_2$ dimer correlation $C_{F_2}^{D}(r)$ and the
$\Gamma$-$E_2$ spin correlation $C_{\Gamma-E_2}^{zz}(r)$ for
$L=6,8,10$ and different separations $r$. The dashed line denotes
$C_{F_2}^{D}=A C_{\Gamma-E_2}^{zz}$ with $A\simeq2.52$.
The inset shows the symmetry-resolved $M$-point dimer modes, with the
$F_2$ sector dominant for all system sizes. }\label{fig2}
\end{figure*}

In Fig.~\ref{fig2}\textbf{a}, we plot the spin structure factor $S^{zz}_{f} (\mathbf{q})$ in an extended Brillouin zone.
We show that no sharp peak develops inside the first Brillouin zone, while the maxima occur at the reciprocal-lattice images $G$ of $\Gamma$. Since $G$ differs from $\Gamma$ only by a reciprocal-lattice vector, the corresponding real-space correlations characterize a zero-Bravais-momentum three-sublattice fluctuation pattern repeated across unit cells, rather than a distinct finite-momentum modulation. The nontrivial structure resides within the unit cell, where the $A$, $B$, and $C$ sublattice components form predominantly the nonuniform $\Gamma-E_2$ sector through the kagome sublattice form factor (see Supplemental Material).
As shown in Fig.~\ref{fig2}\textbf{b}, the normalized peak values of spin structure factors $S^{zz}_{\text{max}}/N$ tend to 0 with $N\rightarrow \infty$, indicating the absence of long-range magnetic order. We also provide the data of $S^{D}_{f} (\mathbf{q})$ and its 
normalized maxima $S^{D}_{\text{max}}$ in the Supplementary Material, showing that $S^{D}_{\text{max}}/N \rightarrow 0$ with increasing $N$, which rules out the valence-bond-solid (VBS) order. Additionally, the dramatically large value of $S^{D}_{\text{max}} \sim  87.37$  for $L=6$ reported by Ref.~\cite{PhysRevX.15.011047} is not reproduced in our calculations. Instead, $S^{D}_{\text{max}}$ systematically decreases with increasing $L$. Thus, we argue that the signatures of an SPDW ground state reported for Hamiltonian~(\ref{eq_h}) in Ref.~\cite{PhysRevX.15.011047} require further validation, in particular through explicit checks of the ergodicity of the Monte Carlo sampling. 

The dimer structure factors (\ref{dimer}) consider a particular bond form factor fixed by the bond-center positions. To resolve the full bond-correlation structure, we then focus on the connected bond-correlation matrix 
\begin{eqnarray}
K_{\alpha,\beta}^{\text{conn}} (\mathbf{q}) = \frac{1}{N} \sum_{\mathbf{R},\mathbf{R'}} e^{i\mathbf{q}\cdot(\mathbf{R} - \mathbf{R'})} \langle \delta B_{\alpha}(\mathbf{R}) \delta B_{\beta}(\mathbf{R'}) \rangle, 
\label{k_conn}
\end{eqnarray}
where $B_{\alpha}(\mathcal{R})$ is the nearest-neighbor bond-energy operator, $\alpha=1,\ldots,6$ labels the complete set of six
nearest-neighbor bonds associated with each Bravais unit cell, and
$\delta B_{\alpha}(\mathbf R)
= B_{\alpha}(\mathbf R)-\langle B_{\alpha}(\mathbf R)\rangle$.
In Fig.~\ref{fig2}\textbf{c}, we plot the largest eigenvalue of the matrix $K_{\alpha,\beta}^{\text{conn}} (\mathbf{q})$, i.e., $\lambda_{1}[K^{\text{conn}} (\mathbf{q})]$, which quantifies the strongest connected dimer fluctuation at the momentum $\mathbf{q}$. It is seen that $\lambda_{1}[K^{\text{conn}} (\mathbf{q})]$ is maximal at the $M$ points. 
In real space, the dominant $M$-point dimer fluctuations correspond to period-doubled bond modulations, with the bond pattern alternating under translations along one or both primitive Bravais-lattice directions. The three $M$ points represent the three symmetry-related orientations of this modulation. Condensation in the $M$-point  dimer channel implies broken lattice-translation symmetry, while 
$\lambda_{1}[K^{\text{conn}} (M)]/N \rightarrow 0$ with increasing $N$ precludes long-range VBS order (see Fig.~\ref{fig2}\textbf{b}).

Next, we resolve the $M$-point dimer fluctuations into
symmetry-adapted sectors of the kagome space group, including
$F_{2}$, $B_{2}$, and $F_{1}$ (see Supplemental Material for their
explicit construction and symmetry projectors). As shown in the inset
of Fig.~\ref{fig2}\textbf{d}, the $F_{2}$ sector is dominant for all
$L$. Together with Fig.~\ref{fig2}\textbf{a}, this identifies the
$M$-$F_{2}$ dimer and $\Gamma$-$E_{2}$ spin channels as the dominant
symmetry-resolved fluctuations. 
To further examine the relation between these two channels, we compare the real-space
correlations in the aforementioned two dominant symmetry channels. We define
the sublattice-resolved spin and bond correlations as
\begin{eqnarray}
C_{\alpha\beta}^{zz}(\mathbf r)
&=&
\frac{1}{N}\sum_{\mathbf R}
\left\langle
S_{\alpha}^{z}(\mathbf R)
S_{\beta}^{z}(\mathbf R+\mathbf r)
\right\rangle ,
\nonumber\\
C_{\alpha\beta}^{D}(\mathbf r)
&=&
\frac{1}{N}\sum_{\mathbf R}
\left\langle
\delta B_{\alpha}(\mathbf R)
\delta B_{\beta}(\mathbf R+\mathbf r)
\right\rangle ,
\label{eq_real_corr} 
\end{eqnarray}
where $\alpha,\beta=A,B,C$ for the spin correlations, while
$\alpha,\beta=1,\ldots,6$ label the nearest-neighbor bonds for the
dimer correlations defined above. The $\Gamma$-$E_{2}$ spin correlation is obtained by projecting the
three-sublattice correlation matrix onto the $E_{2}$ sector, i.e., $C_{\Gamma-E_{2}}^{zz}(\mathbf r)
=
\frac{1}{2}
\sum_{\alpha,\beta=A,B,C}
(P_{E_{2}})_{\alpha\beta}
C_{\alpha\beta}^{zz}(\mathbf r)$ with $P_{E_{2}} = I-\mathbf u\mathbf u^{\dagger}$ and $\mathbf u=\frac{1}{\sqrt{3}}(1,1,1)^{T}$.  Similarly, we define $C_{F_{2}}^{D}(\mathbf r)
=
\frac{1}{3}
\sum_{i=1}^{3}
e^{-i\mathbf M_i\cdot\mathbf r}
\left[\mathbf v_{F_{2}}^{(i)}\right]^{\dagger}
C^{D}(\mathbf r)
\mathbf v_{F_{2}}^{(i)}$, where the explicit $F_{2}$ bond vectors and their symmetry
construction are given in the Supplemental Material.

In the $N_f=4$ QED$_3$ description of the $U(1)$ Dirac QSL of the kagome Heisenberg antiferromagnet, the
$M$-$F_{2}$ dimer and $\Gamma$-$E_{2}$ spin channels carry the
quantum numbers of spin-singlet and spin-triplet minimal monopoles,
respectively~\cite{Song:2019wv}. These monopoles are components of
the same emergent $SO(6)$ multiplet and thus share a common scaling
dimension~\cite{Song:2019wv,Seifert:2024vp}. In Fig.~\ref{fig2}\textbf{d}, we directly compare
$C_{F_{2}}^{D}(r)$ and $C_{\Gamma-E_{2}}^{zz}(r)$ for different distances $r$ and system sizes $L$. Remarkably, all data collapse onto a proportional function $C_{F_{2}}^{D}(r)
\simeq
A\,C_{\Gamma-E_{2}}^{zz}(r)$ with $A\simeq2.52$. This is precisely the behavior expected for different components of the same emergent $SO(6)$ monopole multiplet, showing that the symmetry-resolved correlations obtained from our NQS calculations are consistent with the emergent monopole structure of the kagome $U(1)$ Dirac QSL.

\emph{Overlap of ground state with candidate QSLs.}-- Previous studies of projected mean-field VMC for the kagome model (\ref{eq_h}) cannot unambiguously distinguish between the between $Z_{2}$ and the U(1) Dirac spin-liquid states because their variational energies are nearly degenerate~\cite{PhysRevB.84.020407,PhysRevB.102.195106}.  Our NQS is not biased toward either of the competing parton mean-field ansätze, providing an independent and more accurate representation of the ground-state wave function of the Hamiltonian (\ref{eq_h}). This motivates us to assess whether the optimized NQS is closer to the U(1) Dirac or $Z_{2}$ spin-liquid manifold. 


For this assessment, we use the following procedure. We determine 
the fidelity between our optimized NQS $|\Psi_{\boldsymbol{\theta}}^{\textrm{opt}}\rangle$ with representative Gutzwiller-projected fermionic states from these competing variational families $|\Phi_{\mu,\nu} \rangle$ with $\mu$ and $\nu$ labeling the variational family and spinon boundary sector, respectively.  The fidelity is given by
\begin{eqnarray}
F_{\mu,\nu} = \frac{|\langle \Psi_{\boldsymbol{\theta}}^{\textrm{opt}} |\Phi_{\mu,\nu}  \rangle |^{2}}{\langle\Psi_{\boldsymbol{\theta}}^{\textrm{opt}}  |\Psi_{\boldsymbol{\theta}}^{\textrm{opt}}   \rangle \langle \Phi_{\mu,\nu} | \Phi_{\mu,\nu} \rangle },
\label{eq_f}
\end{eqnarray}
and can be evaluated by Monte Carlo sampling configurations from the NQS probability distribution. We adopt approximately $3\times 10^{5}$ Monte Carlo samples to obtain fidelity estimates with small statistical uncertainties. As the Dirac spin-liquid reference ($\mu = \textrm{Dirac}$),  we employ the optimized extended $U(1)$ Dirac state in Ref.~\cite{PhysRevB.84.020407}, as one of the lowest projected-mean-field variational energies reported for the Dirac state. For $Z_{2}$ candidates, we consider two complementary references. The first ($\mu = Z_{2}$) is obtained by introducing a finite perturbation away from the Dirac state considered in Ref.~\cite{PhysRevB.84.020407}, without producing any detectable change in the variational energy. The second ($\mu = Z_{2}^{*}$) is the fully optimized $Z_{2}$ resonating-valence-bond state in Ref.~\cite{PhysRevB.102.195106}, 
whose variational energy lies slightly below that of their optimized $U(1)$ state. Following the boundary conditions used in the original VMC studies, we restrict $\nu$ to the two mixed periodic-antiperiodic spinon sectors. For the Dirac state, only nondegenerate closed-shell sectors are considered, however for the $Z_{2}$ states, both mixed sectors are retained. Finally, we compare the maximum fidelity $F_{\mu}^{\textrm{max}}$ over all admissible boundary sectors for different $\mu$. 
More details of the parametrizations and boundary conditions for the chosen mean-field states are given in the Supplemental Material. 

\begin{figure}[]
	\centering
	\includegraphics[width=1\linewidth]{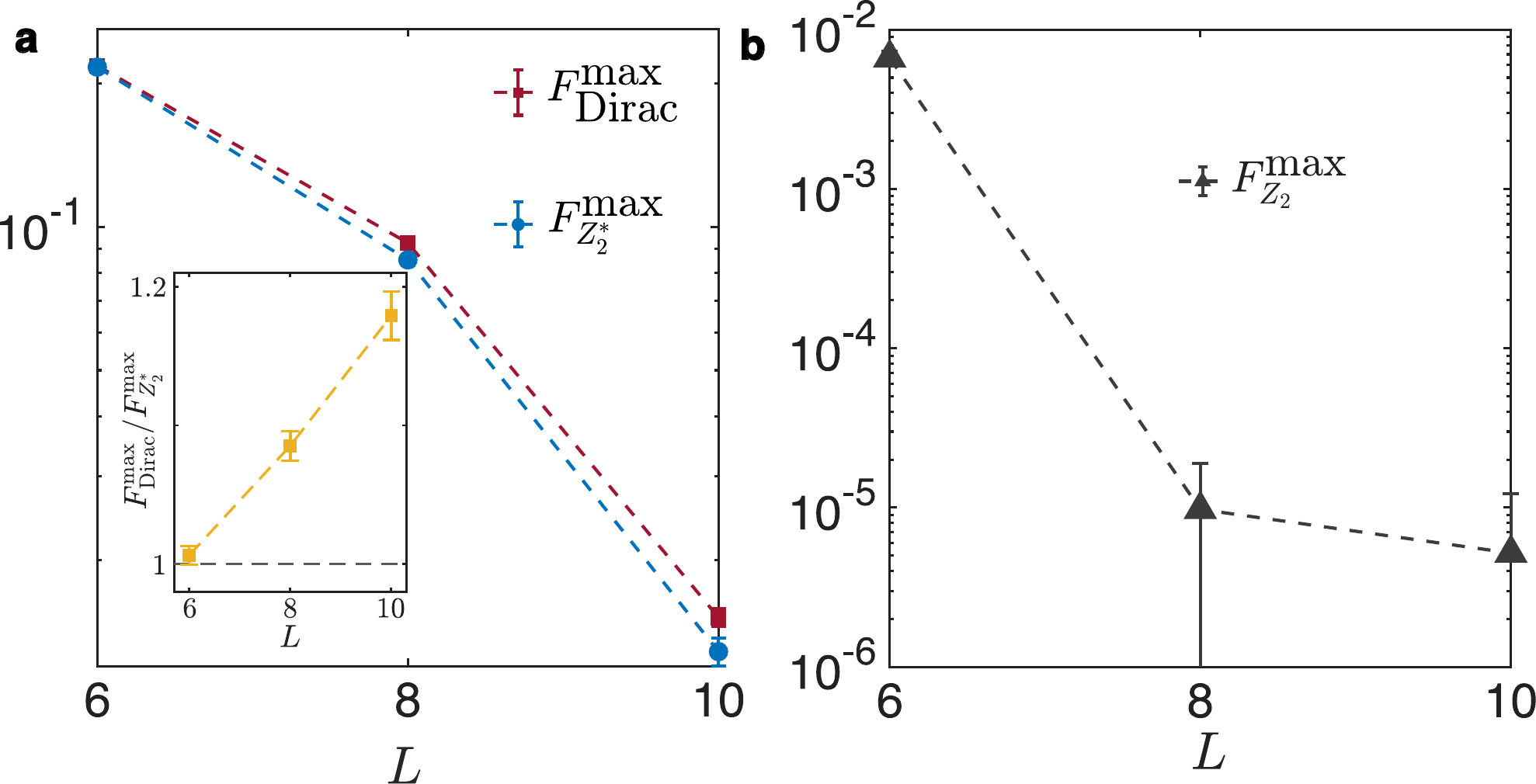}\\
	\caption{\textbf{Fidelity with Gutzwiller-projected spin-liquid reference states.} \textbf{a}, The maximum fidelity $F_{\mu}^{\textrm{max}}$ between the optimized KagomeResNet NQS with $\mu = \textrm{Dirac}$ and $\mu = Z_{2}^{*}$ as a function of system size $L$. The inset shows the ratio between the two fidelities $F^{\textrm{max}}_{\textrm{Dirac}}/F^{\textrm{max}}_{Z_{2}^{*}}$ as a function of $L$.  \textbf{b}, The maximum fidelity $F_{\mu}^{\textrm{max}}$ with $\mu = Z_{2}$. Error bars indicate Monte Carlo statistical uncertainties. Dashed lines connecting data points are guides to the eye. }\label{fig3}
\end{figure}

The results of $F^{\textrm{max}}_{\mu}$ for the aforementioned three Gutzwiller-projected mean-field states and the optimized KagomeResNet NQS with $n=c=40$ are displayed in Fig.~\ref{fig3}. Although the variational energies for the Dirac and $Z_{2}$ ($\mu = \textrm{Dirac}$ and $Z_{2}$) reference states in Ref.~\cite{PhysRevB.84.020407} are indistinguishable, 
$F_{Z_{2}}^{\textrm{max}}$ is significantly smaller than $F_{\textrm{Dirac}}^{\textrm{max}}$. Moreover, for a more competitive $Z_{2}$ reference ($\mu = Z_{2}^{*}$) in Ref.~\cite{PhysRevB.102.195106} with lower variational energy,  $F_{Z_{2}^{*}}^{\textrm{max}}$ is close to $F_{\textrm{Dirac}}^{\textrm{max}}$ at $L=6$, while it becomes systematically smaller than $F_{\textrm{Dirac}}^{\textrm{max}}$ with increasing $L$. The relative separation is highlighted by the ratio $F^{\textrm{max}}_{\textrm{Dirac}}/F^{\textrm{max}}_{Z_{2}^{*}}$, which increases linearly with $L$ (see the inset of Fig.~\ref{fig3}\textbf{a}).

We emphasize that although both the $Z_2$ and $Z_2^{*}$ reference states originate from gapped $Z_2$ mean-field ansätze, the latter, after Gutzwiller projection, exhibits properties resembling those of a Dirac QSL, such as a gapless spin excitation spectrum~\cite{PhysRevB.102.195106}. The much larger fidelity $F_{Z_2^{*}}^{\mathrm{max}}\gg F_{Z_2}^{\mathrm{max}}$ shows that the optimized NQS is substantially closer to this reference with Dirac-like spin properties. Combined with the direct fidelity preference for the Dirac reference at $L=8$ and $10$ (Fig.~\ref{fig3}\textbf{a}), these results support a Dirac spin-liquid interpretation of the NQS ground state.


\emph{Discussion.} -- We have studied the ground-state properties of spin-1/2 kagome Heisenberg antiferromagnet with nearest-neighbor interactions using the KagomeResNet, an NQS architecture developed in this work, featuring message-passing convolutions tailored to the geometry of kagome lattice. In stark contrast to the results in Ref.~\cite{PhysRevX.15.011047}, we show that the ground state lies in the $\Gamma$ instead of $M$ point and no evidence of SPDW states can be identified. The analysis of symmetry-resolved spin and dimer correlations and the wave-function fidelity comparisons support the $U(1)$ Dirac QSL nature for the ground state of the kagome model (\ref{eq_h}).  

Future work can extend the application of KagomeResNet to explore the ground-state properties of other kagome systems, such as the long-range $XY$ kagome model, which can be realized by dipolar Rydberg atoms~\cite{Yao:2018tn,bintz2024diracspinliquidquantum,bornet2026diracspinliquidcandidate}. In addition, we have demonstrated that the KagomeResNet outperforms the conventional ResNet, suggesting that the geometry-aware message-passing convolutions can play a key role in the NQS architecture. This can motivate the design of NQS for solving the ground states of models with more complex geometry, including the spin-1/2 Heisenberg antiferromagnet on the shuriken lattice~\cite{PhysRevB.104.L220408,PhysRevB.107.064406} and 
maple-leaf lattice~\cite{ebert2026competingparamagneticphasesmapleleaf}.

\begin{acknowledgments}
The NQS simulations are performed using
\texttt{\textbf{Quantax}}~\cite{quantax} based on \texttt{\textbf{JAX}}~\cite{jax2018github} and \texttt{\textbf{Equinox}}~\cite{kidger2021equinoxneuralnetworksjax}. We gratefully acknowledge the scientific support and high-performance computing
resources provided by the Erlangen National High Performance Computing Center (NHR) of the Friedrich-Alexander-Universität
Erlangen-Nürnberg (NHR Project No. nqsQuMat), and by the Jülich
Booster (JUPITER Booster) at the Jülich Supercomputing Centre, Forschungszentrum Jülich (project neuquass). This work was supported by the German Research Foundation (DFG) via project 492547816 (TRR 360). C.R. acknowledges support from the Stewart Blusson Quantum Matter Institute at the University of British Columbia.

\end{acknowledgments}


\bibliography{reference_ka}

\end{document}


\title{Supplementary Materials for ``Evidence for a Dirac Spin Liquid in the Kagome Heisenberg Antiferromagnet from High-Precision Neural-Network Quantum States"}

\author{Zheng-Hang Sun}
\affiliation{Theoretical Physics \uppercase\expandafter{\romannumeral3}, Center for Electronic Correlations and Magnetism, Institute of Physics, University of Augsburg, 86135 Augsburg, Germany}

\author{Christopher Roth}
\affiliation{Center for Computational Quantum Physics, Flatiron Institute, 162 5th Avenue, New York, NY 10010, USA}

\author{Ao Chen}
\affiliation{Division of Chemistry and Chemical Engineering, California Institute of Technology, Pasadena, California 91125, USA}

\author{Markus Heyl}
\affiliation{Theoretical Physics \uppercase\expandafter{\romannumeral3}, Center for Electronic Correlations and Magnetism, Institute of Physics, University of Augsburg, 86135 Augsburg, Germany}
\affiliation{Centre for Advanced Analytics and Predictive Sciences (CAAPS),
University of Augsburg, Universitätssr. 12a, 86135 Augsburg, Germany}

\maketitle

\section{Details of network architecture and results for conventional residual convolutional network}

In this section, we present the details of the conventional residual convolutional network ResNet, which is employed 
as a baseline for comparison with KagomeResNet considered in the main text.   

\begin{figure}[]
	\centering
	\includegraphics[width=1\linewidth]{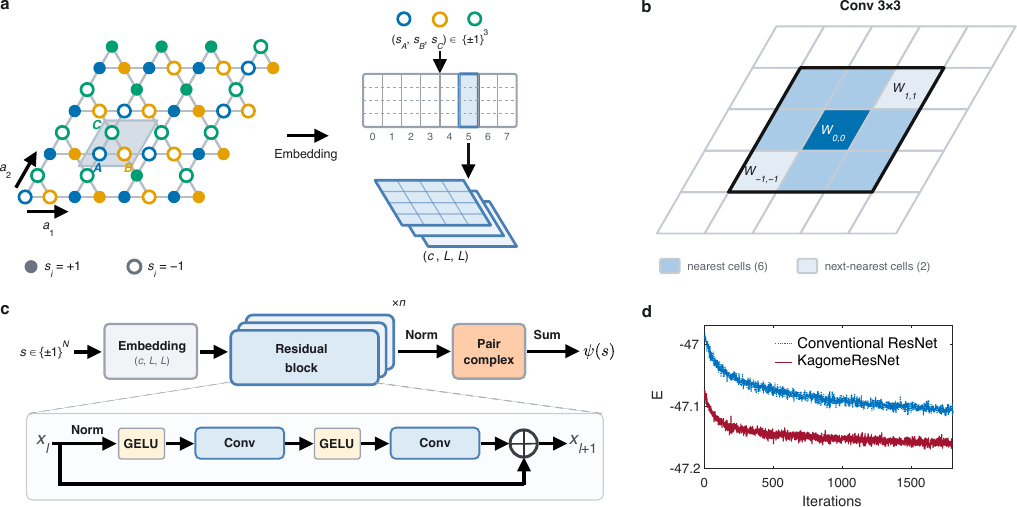}\\
	\caption{\textbf{Conventional residual convolutional network (ResNet) applied directly to the kagome model.} \textbf{a}, Input embedding. The kagome lattice is represented as a triangular Bravais lattice with three sites $A$, $B$, and $C$ in each unit cell. For a spin configuration $s\in \{ \pm 1 \}^{N}$, the three spins $(s_{A},s_{B},s_{C}) \in \{ \pm 1 \}^{3}$ within each unit cell (shaded) are encoded as one of 8 local states. A trainable embedding table maps each local state to a $c$-dimensional feature vector, producing a feature tensor of shape $(c,L,L)$, where $c$ is the number of feature channels. The three-sublattice structure of the kagome lattice is therefore absorbed into the local embedding. \textbf{b}, Conventional $3\times3$ convolution. The subsequent convolutional layers operate on the Bravais-cell indices using a standard \(3\times3\) stencil with periodic (circular) padding. Each of the nine offsets $\delta\in\{-1,0,1\}^{2}$ is associated with a weight matrix $W_{\delta}\in\mathbb{R}^{c \times c}$, including the central weight $W_{0,0}$. On the Bravais lattice, the stencil covers the six nearest-neighbour cells and two additional next-nearest-neighbour cells at $\delta=\pm(1,1)$. Thus, although the local three-spin configuration is retained through the embedding, the convolution itself acts on the Bravais-cell grid and does not explicitly encode the microscopic nearest-neighbour connectivity of the kagome lattice.  \textbf{c}, Network architecture of ResNet. The embedded input is processed by $n$ residual blocks, each consisting of normalization (Norm), GELU activations, and two standard $3\times3$ convolution layers connected by a residual shortcut. After a final normalization, the pair-complex activation combines real channels into complex outputs, which are summed to yield the wave-function amplitude $\psi(\mathbf{s})$.   \textbf{d}, Variational-energy optimization with ResNet for $L=6$ in the $\Gamma$-point sector. The energy trajectory of the conventional ResNet is shown over the final 1,800 optimization iterations, with both calculations imposing \(\mathcal{I}_{s}=+1\) and the \(D_{6}\) point-group symmetry in comparison with the KagomeResNet studied in the main text.  }\label{figs1}
\end{figure}

As shown in Fig.~\ref{figs1}\textbf{a}, the kagome lattice is regarded as a Bravais lattice with three sublattice sites $A$, $B$, and $C$ in each unit cell. 
For an $L\times L$ system, the total number of spins is $N=3L^2$, and a spin configuration can be written as $\mathbf{s} =  \{s_{\alpha}(i,j)\}$ with $\alpha=A,B,C,$, $s_{\alpha}(i,j)=\pm1$, and $(i,j)$ labelling the unit cell along the two primitive vectors $\mathbf{a}_1$ and $\mathbf{a}_2$. The conventional ResNet treats the three spins within each unit cell as a single local degree of freedom. 
Specifically, we first introduce the binary variables $b_{\alpha}(i,j) = \frac{s_{\alpha}(i,j)+1}{2} \in \{0,1\}$ and encode the local spin configuration by the integer 
\begin{equation}
    k(i,j)
    =
    b_A(i,j)
    +2b_B(i,j)
    +4b_C(i,j),
    \label{eq:resconv_embedding_index} 
\end{equation}
with     $k(i,j)\in\{0,\ldots,7\}$. Therefore, the $2^3=8$ possible spin configurations of a kagome unit cell are represented as eight local states.


A trainable embedding matrix $E\in\mathbb{R}^{c\times 8}$ then maps each local state to a $c$-dimensional feature vector, 
\begin{equation}
    X^{(0)}_{\mu,i,j}
    =
    E_{\mu,k(i,j)},
    \qquad
    \mu=1,\ldots,c.
    \label{eq:resconv_embedding}
\end{equation}
The resulting input to the convolutional network therefore has shape $ X^{(0)}\in\mathbb{R}^{c\times L\times L}$. After the embedding layer, all spatial transformations are performed by standard two-dimensional convolutions on the Bravais-cell grid, as illustrated in Fig.~\ref{figs1}\textbf{b}.
For a hidden feature tensor $X\in\mathbb{R}^{c\times L\times L}$, a convolutional layer can be written as
\begin{equation}
    [\mathcal{C}(X)]_{\mu,i,j}
    =
    b_{\mu}
    +
    \sum_{\nu=1}^{c}
    \sum_{\delta_1,\delta_2=-1}^{1}
    W^{(\delta_1,\delta_2)}_{\mu\nu}
    X_{\nu,i+\delta_1,j+\delta_2},
    \label{eq:resconv_standard_conv}
\end{equation}
where
\begin{equation}
    W^{(\delta_1,\delta_2)}
    \in\mathbb{R}^{c\times c}
\end{equation}
is the weight matrix associated with the cell displacement
$\boldsymbol{\delta}=(\delta_1,\delta_2)$.
For periodic boundary conditions, the cell indices in Eq.~(\ref{eq:resconv_standard_conv}) are understood modulo $L$, corresponding to the circular padding used in the implementation. The $3\times3$ stencil contains nine Bravais-cell positions, including the central cell $\boldsymbol{\delta}=(0,0)$. 
On the Bravais lattice, six of the eight surrounding cells, $(\pm1,0)$, $(0,\pm1)$, $(1,-1)$, and $(-1,1)$ are nearest-neighbour Bravais cells, whereas $(1,1)$ and $(-1,-1)$ are next-nearest-neighbour Bravais cells. Hence, the standard $3\times3$ convolution treats these cell-index displacements as independent learnable kernel components.  

Figure~\ref{figs1}\textbf{c} shows the entire ResNet architecture. Starting from the embedded tensor $X_0=X^{(0)}$, the network contains $n$ residual blocks. 
Each block consists of two standard convolutions with GELU nonlinearities and an identity shortcut. 
For the $l$th residual block ($l=0,\ldots,n-1$), the transformation implemented in our calculations is
\begin{align}
    \widetilde{X}_{l}
    &=
    \frac{X_l}{\sqrt{l+1}},
    \\
    Y_l
    &=
    \mathrm{Conv}^{(1)}_l
    \left[
        \mathrm{GELU}(\widetilde{X}_{l})
    \right],
    \\
    Z_l
    &=
    \mathrm{Conv}^{(2)}_l
    \left[
        \mathrm{GELU}(Y_l)
    \right],
    \\
    X_{l+1}
    &=
    X_l+Z_l .
    \label{eq:resconv_block}
\end{align}
Here, the ``Norm'' operation shown in Fig.~\ref{figs1}\textbf{c} is therefore a deterministic rescaling by $1/\sqrt{l+1}$. 
All residual blocks preserve both the number of feature channels $c$ and the spatial dimensions $L\times L$. Both convolutions in each residual block contain trainable biases, except for the second convolution in the final residual block, for which the bias is omitted.

After the final residual block, the feature tensor is rescaled once more,
\begin{equation}
    X_n\rightarrow \frac{X_n}{\sqrt{n+1}}.
    \label{eq:resconv_final_norm}
\end{equation}
The variational parameters of the convolutional network are real. 
When a complex-valued wave function is required, the $c$ real output channels are divided into two equal sets and combined pairwise according to
\begin{equation}
    Z_{\mu,i,j}
    =
    X_{\mu,i,j}
    +i X_{\mu+c/2,i,j},
    \qquad
    \mu=1,\ldots,c/2,
    \label{eq:pair_complex}
\end{equation}
which corresponds to the ``Pair complex'' operation in Fig.~\ref{figs1}\textbf{c}. 
We therefore choose an even number of channels $c$ whenever a complex output is used. 
The complex features are subsequently passed through the default exponential output activation,
\begin{equation}
    Z_{\mu,i,j}\rightarrow
    \exp\!\left(Z_{\mu,i,j}\right),
\end{equation}
which is evaluated in the code using a numerically stabilized scaled representation to avoid overflow. 
Finally, the output is aggregated over the feature channels and symmetrized with respect to lattice translations. 
For the default translationally invariant sector, the final layer implements the zero-momentum projection, yielding the variational wave-function amplitude  $\psi(\mathbf{s})$.

We note that the KagomeResNet considered in the main text follows the same overall residual architecture as the conventional ResNet shown in Fig.~\ref{figs1}\textbf{c}. The two differences lie in the input representation and the convolution operator. In the following, we provide more details of KagomeResNet. 

The first difference from the conventional ResNet concerns the representation of the spin configuration. 
As discussed above, the conventional ResNet groups the three spins within each kagome unit cell into one of eight local states and subsequently maps this state to a $c$-dimensional trainable embedding. 
In contrast, as shown in the Fig.~1\textbf{a} of the main text, KagomeResNet does not employ such an embedding layer. 
Instead, the microscopic kagome structure is retained explicitly throughout the network.
For a kagome lattice with $N=3L^2$ spins, the input spin configuration $ \mathbf{s}
    =
    \{s_{\alpha}(i,j)\}$ with $\alpha=A,B,C,$
is directly reshaped according to
\begin{equation}
    \mathbf{s}\in\mathbb{R}^{3L^2}
    \longrightarrow
    X^{(0)}
    \in
    \mathbb{R}^{1\times3\times L\times L},
    \label{eq:kagome_reshape}
\end{equation}
with $X^{(0)}_{1,\alpha,i,j}= s_{\alpha}(i,j)$. 
Here, the first dimension denotes the feature-channel index, while the second dimension explicitly labels the three kagome sublattices $A$, $B$, and $C$. 
The remaining two indices $(i,j)$ specify the Bravais unit cell. 
The reshape operation contains no trainable parameters and preserves the original spin variables $s_{\alpha}(i,j)=\pm1$ without coarse graining the three sites of a unit cell into a single local state.
Consequently, whereas the conventional ResNet represents the kagome lattice by a tensor of shape $(c,L,L)$ after the embedding layer, KagomeResNet starts from a tensor of shape $(1,3,L,L)$, 
and retains the explicit sublattice dimension throughout all convolutional layers. 
The feature-channel dimension is expanded from $1$ to $c$ only by the first kagome convolution.

The second, and more important, difference is the convolution operator. As shown in the Fig.~1\textbf{b} of the main text, 
instead of applying a standard $3\times3$ convolution on the Bravais-cell grid, KagomeResNet employs a geometry-aware convolution that performs message passing directly along the microscopic nearest-neighbour bonds of the kagome lattice. 
For an input tensor $X\in \mathbb{R}^{c_{\mathrm{in}}\times3\times L\times L}$, 
the kagome convolution produces $ Y\in \mathbb{R}^{c_{\mathrm{out}}\times3\times L\times L}$, 
while preserving the explicit $A$, $B$, and $C$ sublattice axis.
The nearest-neighbour connectivity is encoded by a fixed set of directed bond classes. 
Using the sublattice labels $(A,B,C)\equiv(0,1,2)$ and the Bravais-cell displacement
$\boldsymbol{\delta}=(\delta_1,\delta_2)$, the six undirected nearest-neighbour bond classes used in the implementation are
\begin{equation}
\begin{split}
    &(A,B,-1,0),\qquad (A,B,0,0),\\
    &(A,C,0,-1),\qquad (A,C,0,0),\\
    &(B,C,0,0),\qquad (B,C,1,-1).
\end{split}
\label{eq:kagome_bonds}
\end{equation}
Each tuple $(\alpha,\beta,\delta_1,\delta_2)$ denotes a bond connecting sublattice $\alpha$ in one unit cell to sublattice $\beta$ in a unit cell displaced by
$\boldsymbol{\delta}=\delta_1\mathbf{a}_1+\delta_2\mathbf{a}_2$.
Both directions of every bond are included in the message-passing operation, resulting in twelve directed nearest-neighbour channels per unit cell. For every directed bond $e=(\alpha\rightarrow\beta,\boldsymbol{\delta})$, 
KagomeResNet associates a learnable matrix
\begin{equation}
    W_e
    \in
    \mathbb{R}^{c_{\mathrm{out}}\times c_{\mathrm{in}}},
\end{equation}
which transforms the feature vector on the source site before it is accumulated on the target site. 

For a directed bond class
$e=(\alpha,\beta,\boldsymbol{\delta})$, with
$\boldsymbol{\delta}=\delta_1\mathbf{a}_1+\delta_2\mathbf{a}_2$,
the contribution from sublattice $\alpha$ in unit cell $\mathbf{r}$ to
sublattice $\beta$ in unit cell $\mathbf{r}+\boldsymbol{\delta}$ is defined as
\begin{equation}
    \Delta Y^{(e)}_{\mu,\beta}
    (\mathbf{r}+\boldsymbol{\delta})
    =
    \sum_{\nu=1}^{c_{\mathrm{in}}}
    (W_e)_{\mu\nu}
    X_{\nu,\alpha}(\mathbf{r}),
    \label{eq:kagome_message}
\end{equation}
where $W_e\in\mathbb{R}^{c_{\mathrm{out}}\times c_{\mathrm{in}}}$ is the trainable
weight matrix associated with the directed bond class $e$.
The output feature on each target site is obtained by summing the contributions
from all directed nearest-neighbour bond classes terminating at that site,
together with the corresponding on-site contribution.
The same matrix $W_e$ is shared over all lattice translations of a given
directed bond class, thereby preserving translational equivariance while
explicitly encoding the microscopic connectivity of the kagome lattice.


In Fig.~\ref{figs1}\textbf{d}, we directly compare the energy trajectory of the final 1,800 optimization iterations after imposing \(\mathcal{I}_{s}=+1\) and the \(D_{6}\) point-group symmetry for both the conventional ResNet and KagomeResNet. Here, we consider the conventional ResNet with $n=32$, $c=42$ and $3\times 3$ kernel, whose number of trainable parameters is $N_{\mathrm{para}} = 1019046$, which is slightly larger than $N_{\mathrm{para}} = 974208$ for the KagomeResNet with $n=c=32$ considered in our main text. It is seen that KagomeResNet  outperforms conventional ResNet.

\section{Details of variational optimization}

For both the conventional ResNet and the KagomeResNet, we supplement the complex network output $\psi_{\boldsymbol{\theta}}(\mathbf s)$ with a fixed three-sublattice phase factor associated with a coplanar pattern,
\begin{equation}
 \phi_{q=0}(s)
 =\exp\left[
 i\sum_{j=1}^{N}\frac{1-s_j}{2}\,\varphi_{\alpha(j)}
 \right],
 \qquad
 (\varphi_A,\varphi_B,\varphi_C)
 =\left(0,\frac{2\pi}{3},\frac{4\pi}{3}\right),
 \label{eq:training_reference_phase}
\end{equation}
where $s_j=\pm1$ and $\alpha(j)\in\{A,B,C\}$ labels the sublattice of
site $j$. Each down spin contributes the phase associated with its
sublattice. This factor has unit modulus and is held fixed throughout
training. The complex-valued network $\psi_{\boldsymbol{\theta}}(s)$
can learn additional configuration-dependent phase corrections. Thus,
the reference phase provides a prior without fixing the full phase
structure of the variational state. 

We initially optimize the neural-network quantum state (NQS) with translational invariance imposed. After $10^{4}$ optimization steps, we additionally impose spin-flip symmetry and the spatial symmetries appropriate to the target symmetry sector, and continue training from the previously optimized network parameters. For the optimization method, we adopt AdamSR in Quantax, which combines MinSR~\cite{Chen:2024tb} with Adam-type momentum and adaptive rescaling. We adopt the training rate is $10^{-3}$. 


\section{Additional data for the spin and dimer structure factors}

\begin{figure}[]
	\centering
	\includegraphics[width=1\linewidth]{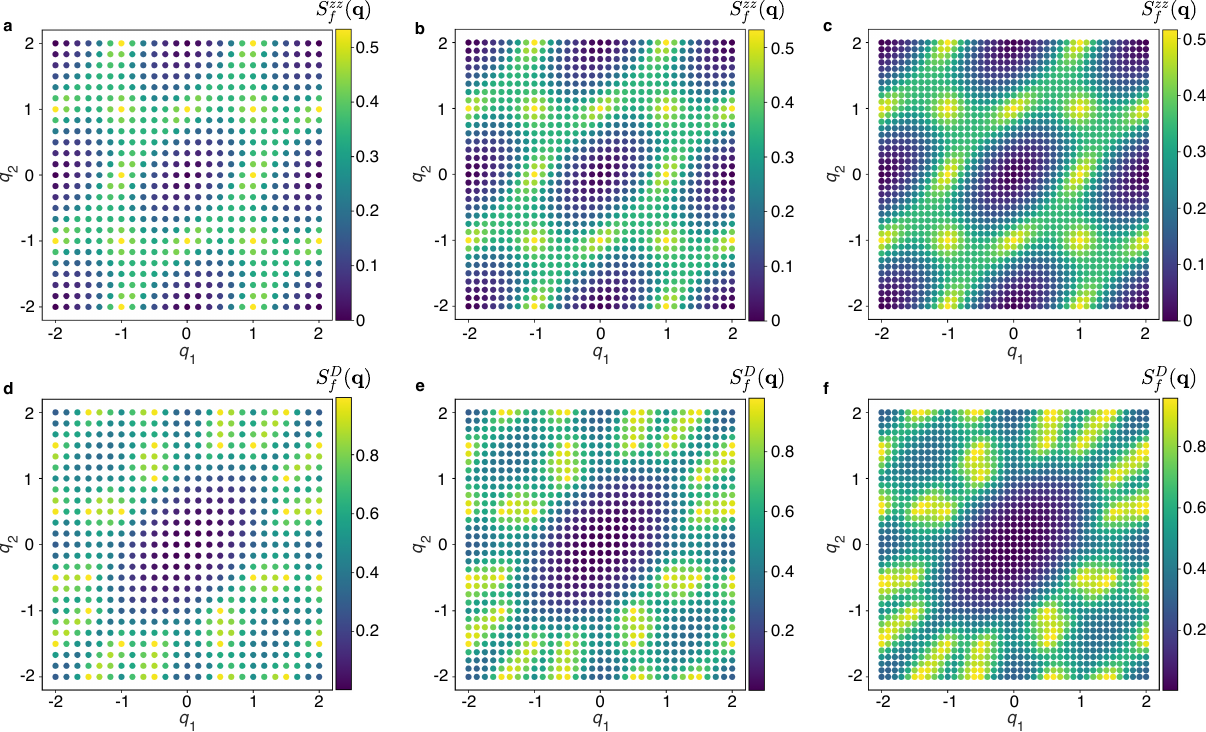}\\
	\caption{\textbf{Spin and dimer structure factors in reciprocal-lattice coordinates.} \textbf{a}-\textbf{c}, Static spin structure factor $S_f^{zz}(\mathbf q)$ for $L=6$, $8$, and $10$, respectively.  \textbf{d}-\textbf{f}, Dimer structure factor $S_f^{D}(\mathbf q)$ ffor $L=6$, $8$, and $10$, respectively. The momentum is expressed in reciprocal-lattice coordinates $(q_{1},q_{2})$.   }\label{figs2}
\end{figure}

In the main text, we present the spin structure factor for $L=10$ as a
function of the wave vector $\mathbf{q}=(q_x,q_y)$ expressed in Cartesian
reciprocal-space coordinates (see Fig.~2\textbf{a} of the main text).
For a direct comparison with Ref.~\cite{PhysRevX.15.011047}, we additionally present
the structure factors using the reciprocal-lattice coordinates $(q_1,q_2)$
employed therein. We emphasize that $q_1$ and $q_2$ are not Cartesian
components of the momentum, but dimensionless coefficients with respect to
the two primitive reciprocal-lattice vectors. 

Using the real-space primitive vectors $\mathbf{a}_1=(1,0)$, $ \mathbf{a}_2=\left(\frac{1}{2},\frac{\sqrt{3}}{2}\right)$, 
the corresponding reciprocal-lattice vectors are $\mathbf{b}_1=2\pi\left(1,-\frac{1}{\sqrt{3}}\right)$, $ \mathbf{b}_2 = 2\pi\left(0,\frac{2}{\sqrt{3}}\right)$, 
which satisfy $\mathbf{a}_i\cdot\mathbf{b}_j=2\pi\delta_{ij}$.
The two momentum representations are therefore related by $\mathbf{q} = q_1\mathbf{b}_1+q_2\mathbf{b}_2$. 
For an $L\times L$ torus, the momenta in the first reciprocal unit cell are
quantized as
\begin{equation}
    q_1=\frac{n_1}{L},
    \qquad
    q_2=\frac{n_2}{L},
    \qquad
    n_1,n_2=0,\ldots,L-1.
\end{equation}
Momentum points in an extended-zone representation are obtained by reciprocal
lattice translations,
\begin{equation}
    (q_1,q_2)
    \rightarrow
    (q_1+m_1,q_2+m_2),
    \qquad m_1,m_2\in\mathbb{Z},
\end{equation}
which corresponds in Cartesian coordinates to
$\mathbf{q}\rightarrow\mathbf{q}
+m_1\mathbf{b}_1+m_2\mathbf{b}_2$.

\begin{figure}[]
	\centering
	\includegraphics[width=1\linewidth]{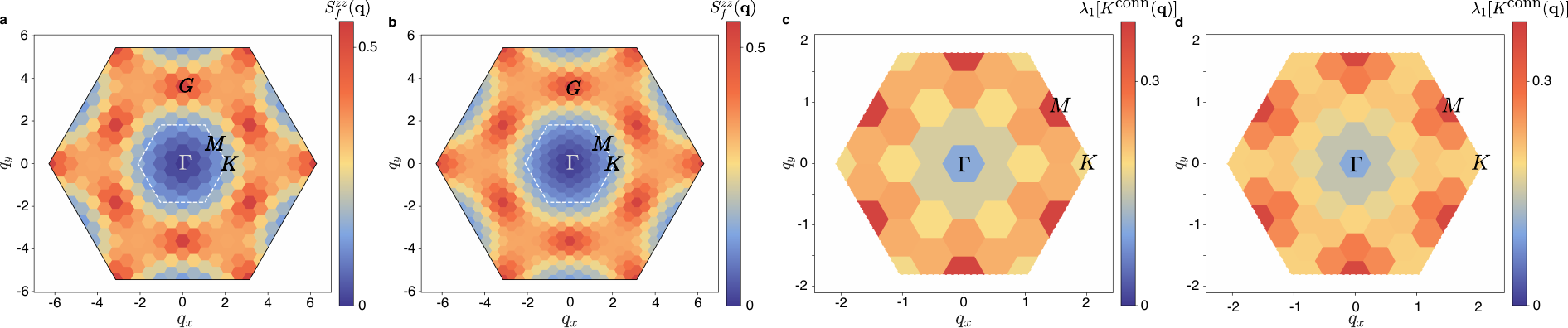}\\
	\caption{\textbf{Additional data of structure factors and largest eigenvalues of  bond-correlation matrix.} \textbf{a}-\textbf{b}, Static spin structure factor $S_f^{zz}(\mathbf q)$ in Cartesian
reciprocal-space coordinates $\mathbf{q}=(q_x,q_y)$ for  $L=6$ and $8$, respectively. \textbf{c}-\textbf{d}, Largest eigenvalue of the connected bond-correction matrix $\lambda_{1} [K^{\textrm{conn}}(\mathbf q)]$ in the coordinates $\mathbf{q}=(q_x,q_y)$ for  $L=6$ and $8$, respectively.  }\label{figs3}
\end{figure}

Figure~\ref{figs2} shows the corresponding spin and dimer structure factors
for $L=6$, $8$, and $10$ in the $(q_1,q_2)$ representation. For $L=6$, Ref.~\cite{PhysRevX.15.011047} reported a dramatic enhancement
of $S_f^{D}(\mathbf q)$ relative to the $L=4$ result. In particular, while
the maximum value for $L=4$ was approximately $0.933$, the reported
maximum increased to $S_{\max}^{D}\simeq87.37$ for $N=108$, accompanied by
numerous pronounced peaks throughout the extended reciprocal space.
This strong enhancement was interpreted in
Ref.~\cite{PhysRevX.15.011047} as a signature of the proposed spinon
pair-density-wave state and its associated secondary orders. We do not observe such behavior. Already for the same $L=6$ ($N=108$)
torus (Fig.~\ref{figs2}\textbf{d}), the dimer structure factor remains of
order unity and exhibits no anomalously enhanced peaks. Upon increasing
the system size to $L=8$ and $L=10$
(Figs.~\ref{figs2}\textbf{e} and \textbf{f}), no growth analogous to that
reported in Ref.~\cite{PhysRevX.15.011047} develops. Instead, the maximum
value $S_{\max}^{D}$ decreases with increasing $L$, while
$S_{\max}^{D}/N$ extrapolates toward zero, as shown in Fig.~2\textbf{b} of
the main text. The absence of any size-enhanced dimer peak is therefore
inconsistent with the anomalously large dimer response reported in
Ref.~\cite{PhysRevX.15.011047} and provides no support for the associated
SPDW-induced dimer ordering tendency.

In the Fig.~2 of the main text, we present the data of spin structure factors $S_f^{zz}(\mathbf q)$ and largest eigenvalue of the connected bond-correction matrix $\lambda_{1} [K^{\textrm{conn}}(\mathbf q)]$ in Cartesian
reciprocal-space coordinates $\mathbf{q}=(q_x,q_y)$ with $L=10$. Here, in Fig.~\ref{figs3}, we also display the results of $S_f^{zz}(\mathbf q)$ and $\lambda_{1} [K^{\textrm{conn}}(\mathbf q)]$ ($\mathbf{q}=(q_x,q_y)$) with other two system sizes, i.e., $L=6$ and $8$
For all three system sizes, the maxima of $S_f^{zz}(\mathbf q)$ occur at the reciprocal-lattice images $G$ of the $\Gamma$ point, whereas the maxima of $\lambda_{1}[K^{\mathrm{conn}}(\mathbf q)]$ are located at the $M$ points. The locations of these dominant spin and dimer fluctuations are therefore robust against changes in system size.



\section{Symmetry-resolved spin and dimer correlations}
\label{sec:symm_corr}

In this section, we provide the symmetry construction underlying the
$\Gamma$-$E_2$ spin and $M$-$F_2$ dimer correlations discussed in
Fig.~2 of the main text.  The purpose of this analysis is to go beyond
the momentum dependence of the scalar structure factors and resolve the
internal sublattice or bond structure of the dominant fluctuations.
As we show below, the spin maxima at reciprocal-lattice images of
$\Gamma$ predominantly probe the nonuniform $E_2$ sublattice sector,
whereas the dominant dimer fluctuations at the three $M$ points belong
to the $F_2$ bond sector.  These are precisely the symmetry channels
that can be compared with the monopole operator content of the
$N_f=4$ QED$_3$ description of the kagome Dirac spin liquid (DSL). 

\subsection{$\Gamma$-$E_2$ spin channel}

We first consider the spin correlations.  For sublattices
$\alpha,\beta=A,B,C$, we define the $3\times3$ real-space correlation
matrix
\begin{equation}
    C_{\alpha\beta}^{zz}(\mathbf r)
    =
    \frac{1}{N}
    \sum_{\mathbf R}
    \left\langle
        S_{\alpha}^{z}(\mathbf R)
        S_{\beta}^{z}(\mathbf R+\mathbf r)
    \right\rangle ,
    \label{eq:supp_spin_matrix}
\end{equation}
where $\mathbf R$ labels the Bravais cells and
$N=3L^2$.  Correspondingly, the sublattice-resolved momentum-space
correlation matrix is obtained by Fourier transforming
$C_{\alpha\beta}^{zz}(\mathbf r)$.

The physical spin structure factor defined in the main text contains,
in addition, the kagome sublattice form factor.  Denoting the position
of sublattice $\alpha$ inside the unit cell by
$\boldsymbol{\rho}_{\alpha}$, it can be written as
\begin{equation}
    S_f^{zz}(\mathbf q)
    =
    \sum_{\alpha,\beta}
    e^{i\mathbf q\cdot
    (\boldsymbol{\rho}_{\alpha}-\boldsymbol{\rho}_{\beta})}
    \,
    \mathcal S_{\alpha\beta}^{zz}(\overline{\mathbf q}),
    \label{eq:supp_spin_formfactor}
\end{equation}
where $\overline{\mathbf q}$ denotes the momentum folded back into the
first Brillouin zone.  This expression also explains why the physical
structure factor of a non-Bravais lattice is not generally periodic
under $\mathbf q\rightarrow\mathbf q+\mathbf G$: although
$\overline{\mathbf q}$ is unchanged, the sublattice form factors
$e^{i\mathbf G\cdot\boldsymbol{\rho}_{\alpha}}$ need not be uniform. 

At the $\Gamma$ point, the three-dimensional sublattice space
decomposes into the uniform one-dimensional component and its
two-dimensional orthogonal complement.  The normalized uniform vector is
\begin{equation}
    \mathbf u=
    \frac{1}{\sqrt 3}
    \begin{pmatrix}
       1\\1\\1
    \end{pmatrix},
\end{equation}
with projector
\begin{equation}
    P_{A_1}=\mathbf u\mathbf u^{\dagger}.
\end{equation}
The complementary two-dimensional subspace transforms as the $E_2$
representation and is selected by
\begin{equation}
    P_{E_2}
    =
    I-P_{A_1}
    =
    I-\mathbf u\mathbf u^{\dagger}.
    \label{eq:supp_PE2}
\end{equation}
For example, an orthonormal basis of this subspace can be chosen as
\begin{equation}
    \mathbf e_1=
    \frac{1}{\sqrt2}
    \begin{pmatrix}
       1\\-1\\0
    \end{pmatrix},
    \qquad
    \mathbf e_2=
    \frac{1}{\sqrt6}
    \begin{pmatrix}
       1\\1\\-2
    \end{pmatrix},
\end{equation}
such that
$P_{E_2}=\mathbf e_1\mathbf e_1^\dagger+
\mathbf e_2\mathbf e_2^\dagger$.

We therefore define the symmetry-resolved $\Gamma$-$E_2$ correlation
used in the main text as
\begin{equation}
    C_{\Gamma-E_2}^{zz}(\mathbf r)
    =
    \frac{1}{2}
    \sum_{\alpha,\beta=A,B,C}
    (P_{E_2})_{\alpha\beta}
    C_{\alpha\beta}^{zz}(\mathbf r).
    \label{eq:supp_CGammaE2}
\end{equation}
The factor $1/2$ averages over the two components of the $E_2$
representation.  In contrast, contraction with $P_{A_1}$ gives the
uniform $\Gamma$-point spin channel.

Equation~(\ref{eq:supp_spin_formfactor}) provides the connection to
Fig.~2\textbf{a} of the main text.  At a reciprocal-lattice image
$\mathbf q=\mathbf G$ of $\Gamma$, the Bravais momentum is again
$\Gamma$, but the vector of sublattice form factors
\begin{equation}
    \mathbf f(\mathbf G)
    =
    \left(
    e^{i\mathbf G\cdot\boldsymbol{\rho}_A},
    e^{i\mathbf G\cdot\boldsymbol{\rho}_B},
    e^{i\mathbf G\cdot\boldsymbol{\rho}_C}
    \right)^T
\end{equation}
contains a nonuniform component in the $E_2$ subspace.
Consequently, the pronounced $G$-point maxima of
$S_f^{zz}(\mathbf q)$ probe predominantly the nonuniform
$\Gamma$-$E_2$ fluctuations rather than a distinct finite-momentum
magnetic mode.  The persistence of the same momentum and sublattice
structure for $L=6$, $8$, and $10$, shown in Figs.~\ref{figs2} and~\ref{figs3} of the
Supplemental Material and Fig.~2 of the main text, identifies
$\Gamma$-$E_2$ as the dominant symmetry-resolved spin channel.

This identification also has a direct field-theory interpretation.
For the kagome $U(1)$ DSL, Ref.~\cite{Song:2019wv} 
shows that the spin-triplet minimal monopoles carry zero crystal
momentum and angular momentum $l=2$ under the sixfold lattice
rotation.  On the kagome lattice these quantum numbers correspond
precisely to the $\Gamma$-$E_2$ sector.  Thus,
$C_{\Gamma-E_2}^{zz}$ provides a microscopic lattice observable
sensitive to the spin-triplet monopole channel of the DSL.

\subsection{$M$-point bond space and symmetry projectors}

We then resolve the dimer correlations.  Since a kagome unit cell
contains six nearest-neighbour bonds, retaining the complete internal
bond structure requires a six-dimensional bond basis.  We use the
same bond convention in the measurements and throughout the analysis:
\begin{align}
    B_1(\mathbf R)
    &=
    \mathbf S_A(\mathbf R)\cdot\mathbf S_B(\mathbf R),
    \nonumber\\
    B_2(\mathbf R)
    &=
    \mathbf S_A(\mathbf R)\cdot
    \mathbf S_B(\mathbf R-\mathbf a_1),
    \nonumber\\
    B_3(\mathbf R)
    &=
    \mathbf S_A(\mathbf R)\cdot\mathbf S_C(\mathbf R),
    \nonumber\\
    B_4(\mathbf R)
    &=
    \mathbf S_A(\mathbf R)\cdot
    \mathbf S_C(\mathbf R-\mathbf a_2),
    \nonumber\\
    B_5(\mathbf R)
    &=
    \mathbf S_B(\mathbf R)\cdot\mathbf S_C(\mathbf R),
    \nonumber\\
    B_6(\mathbf R)
    &=
    \mathbf S_B(\mathbf R)\cdot
    \mathbf S_C(\mathbf R+\mathbf a_1-\mathbf a_2).
    \label{eq:supp_bond_basis}
\end{align}
We denote
$\delta B_\alpha(\mathbf R)
=B_\alpha(\mathbf R)-\langle B_\alpha\rangle$
and collect these operators into the vector $\delta\mathbf B(\mathbf R)  =
    \big(
    \delta B_1,\ldots,\delta B_6
    \big)^T$. 
The connected real-space bond-correlation matrix is
\begin{equation}
    C_{\alpha\beta}^{D}(\mathbf r)
    =
    \frac{1}{N}
    \sum_{\mathbf R}
    \left\langle
       \delta B_\alpha(\mathbf R)
       \delta B_\beta(\mathbf R+\mathbf r)
    \right\rangle ,
    \label{eq:supp_dimer_matrix}
\end{equation}
whose Fourier transform gives the connected matrix
$K^{\rm conn}_{\alpha\beta}(\mathbf q)$ defined in the main text.

The strongest eigenvalue of $K^{\rm conn}(\mathbf q)$ occurs at the
three symmetry-related $M$ points.  In the reduced reciprocal-lattice
coordinates introduced in Sec.~II, we choose $M_1=\left(\frac12,0\right)$, $M_2=\left(0,\frac12\right)$, and $M_3=\left(\frac12,\frac12\right)$. 

The three momenta form the $M$ star under the $D_6$ point group.

At an individual $M_i$, the little co-group is $C_{2v}$.  We take its
generators to be a twofold rotation $C_2$ about a hexagon center and
a reflection $\tau_i$ whose axis leaves $M_i$ invariant.  A lattice
symmetry acts not only by permuting the six bonds but, whenever a bond
is mapped to a neighbouring reference unit cell, also generates the
corresponding Bloch phase at $M_i$.  We denote the resulting
six-dimensional bond-space representation by $D_{M_i}(g)$.

For a one-dimensional irreducible representation $\lambda$ of the
little group, the symmetry projector is
\begin{equation}
    P_{\lambda}^{(i)}
    =
    \frac{1}{|C_{2v}|}
    \sum_{g\in C_{2v}}
    \chi_{\lambda}^{*}(g)
    D_{M_i}(g),
    \label{eq:supp_group_projector}
\end{equation}
where $\chi_\lambda(g)$ is the corresponding character.
Equivalently, labeling a one-dimensional sector by its eigenvalues
$\eta_2=\pm1$ and $\eta_\tau=\pm1$ under $C_2$ and $\tau_i$,
respectively, Eq.~(\ref{eq:supp_group_projector}) becomes
\begin{equation}
\begin{split}
    P_{\eta_2,\eta_\tau}^{(i)}
    =
    \frac14\big[
       I
       +\eta_2 D_{M_i}(C_2)
       +\eta_\tau D_{M_i}(\tau_i)
       +\eta_2\eta_\tau
        D_{M_i}(C_2\tau_i)
    \big].
    \label{eq:supp_C2v_projector}
\end{split}
\end{equation}
This construction includes the momentum-dependent phases and therefore
classifies the full microscopic bond patterns, rather than only their
orientations inside one unit cell.

For our six-bond basis, the representation at a representative
$M$ point decomposes as
\begin{equation}
    {\cal B}_{M}
    =
    2A_1\oplus A_2\oplus B_1\oplus2B_2.
    \label{eq:supp_bond_decomp}
\end{equation}
In the $M$-star notation used in the main text, the $A_1$ and $A_2$
sectors generate the $F_1$ and $F_2$ bond modes, respectively.
The leading eigenmode of $K^{\rm conn}(M_i)$ belongs to the
one-dimensional, mirror-odd $A_2$ sector, i.e. the $F_2$ sector.
The next modes are classified analogously by applying the projectors
in Eq.~(\ref{eq:supp_C2v_projector}); the dominant sectors displayed
in the inset of Fig.~2\textbf{d} of the main text are $F_2$, $B_2$,
and $F_1$.

\subsection{Explicit $F_2$ bond vectors}

With the bond ordering in Eq.~(\ref{eq:supp_bond_basis}), a normalized
choice of the $F_2$ vectors at the three members of the $M$ star is
\begin{align}
    \mathbf v_{F_2}^{(1)}
    &=
    \frac12
    \begin{pmatrix}
       -1\\
        1\\
        0\\
        0\\
        1\\
        1
    \end{pmatrix},
    &
    M_1&=\left(\frac12,0\right),
    \nonumber\\[2mm]
    \mathbf v_{F_2}^{(2)}
    &=
    \frac12
    \begin{pmatrix}
        0\\
        0\\
        1\\
       -1\\
       -1\\
        1
    \end{pmatrix},
    &
    M_2&=\left(0,\frac12\right),
    \nonumber\\[2mm]
    \mathbf v_{F_2}^{(3)}
    &=
    \frac12
    \begin{pmatrix}
       -1\\
       -1\\
        1\\
        1\\
        0\\
        0
    \end{pmatrix},
    &
    M_3&=\left(\frac12,\frac12\right).
    \label{eq:supp_F2_vectors}
\end{align}
The three vectors are related by the $C_6$ lattice rotations that
permute the members of the $M$ star, including the Bloch phases
associated with bonds crossing the unit-cell boundary.
Their overall signs are conventional and have no physical
consequence, since all projected correlations involve the rank-one
projectors
\begin{equation}
    P_{F_2}^{(i)}
    =
    \mathbf v_{F_2}^{(i)}
    [\mathbf v_{F_2}^{(i)}]^\dagger .
    \label{eq:supp_PF2}
\end{equation}

The $F_2$ contribution to the connected bond covariance at a given
$M_i$ is therefore
\begin{equation}
    \lambda_{F_2}(M_i)
    =
    [\mathbf v_{F_2}^{(i)}]^\dagger
    K^{\rm conn}(M_i)
    \mathbf v_{F_2}^{(i)}.
    \label{eq:supp_lambdaF2}
\end{equation}
Numerically, the leading eigenvector of $K^{\rm conn}(M_i)$ has
essentially unit overlap with this symmetry sector for all three
system sizes.  The ordering of the leading symmetry-resolved
eigenvalues, $F_2>B_2>F_1$, also remains unchanged from
$L=6$ to $L=10$.  Thus, the identification of the dominant
$M$-point fluctuation with $F_2$ is not a consequence of a particular
system size or of an unconstrained numerical eigenvector, but follows
from a fixed symmetry projector.

To compare this finite-momentum dimer fluctuation directly with the
$\Gamma$-point spin channel, we remove the oscillating $M$-point phase
and average over the three members of the $M$ star.  This gives the
real-space correlation used in the main text,
\begin{equation}
    C_{F_2}^{D}(\mathbf r)
    =
    \frac13
    \sum_{i=1}^{3}
    e^{-i\mathbf M_i\cdot\mathbf r}
    [\mathbf v_{F_2}^{(i)}]^\dagger
    C^{D}(\mathbf r)
    \mathbf v_{F_2}^{(i)}.
    \label{eq:supp_CF2}
\end{equation}


The symmetry analysis above identifies the microscopic structure underlying Fig.~2\textbf{c,d} of the main text. The dominant bond fluctuations are centered at the $M$ star and exhibit an internal six-bond pattern that lies overwhelmingly in the mirror-odd $F_2$ sector.

\subsection{Connection to the QED$_3$ monopole multiplet}

Next, we discuss why the two symmetry channels singled out above are
particularly relevant for the Dirac-spin-liquid interpretation.
The low-energy theory of the kagome $U(1)$ DSL is
$N_f=4$ QED$_3$.  The six lowest charge-one monopole operators
transform in the antisymmetric representation $\mathbf 6$ of
SU(4), or equivalently as a vector of the emergent SO(6) symmetry.
They decompose into three spin-singlet and three spin-triplet
components.  Their lattice quantum numbers on the kagome lattice were
clarified in Ref.~\cite{Song:2019wv}.  In particular, the
spin-triplet monopoles have zero crystal momentum and angular momentum
$l=2$ under $C_6$, identifying them with the $\Gamma$-$E_2$ spin
sector considered above.

The spin-singlet monopoles instead carry the translation quantum
numbers of the three $M$ points and correspond to VBS fluctuations.
Matching the translation and little-group transformation laws in the 
Table~3 of Ref.~\cite{Song:2019wv} to the bond-space representation
constructed above identifies the mirror-odd component with the
$M$-$F_2$ sector.  This observation is useful because the symmetry
resolution separates the physical dimer response into operators with
different continuum quantum numbers, which are mixed together in the
scalar dimer structure factor $S_f^D(\mathbf q)$.

A central consequence of the emergent SO(6) structure is that all six
components of the minimal-monopole multiplet have the same scaling
dimension $\Delta_\Phi$.  In the continuum theory their leading
two-point function takes the form
\begin{equation}
    \left\langle
        \Phi_a^\dagger(\mathbf x)
        \Phi_b(\mathbf 0)
    \right\rangle
    \propto
    \frac{\delta_{ab}}
         {|\mathbf x|^{2\Delta_\Phi}},
    \qquad
    a,b=1,\ldots,6,
    \label{eq:supp_SO6_corr}
\end{equation}
as discussed in Refs.~\cite{Song:2019wv,Seifert:2024vp}.
The microscopic spin and dimer operators generally have different
nonuniversal overlaps with their corresponding continuum monopole
operators.  Consequently, their long-distance correlations need not
have the same amplitude, but they are expected to share the same
scaling dependence,
\begin{align}
    C_{\Gamma-E_2}^{zz}(r)
    &\simeq
    A_s\,r^{-2\Delta_\Phi},
    \nonumber\\
    C_{F_2}^{D}(r)
    &\simeq
    A_d\,r^{-2\Delta_\Phi},
    \label{eq:supp_common_scaling}
\end{align}
up to subleading lattice and finite-size corrections.  It follows that
in the regime controlled by the common monopole operator,
\begin{equation}
    C_{F_2}^{D}(r)
    \simeq
    A\,C_{\Gamma-E_2}^{zz}(r),
    \qquad
    A=\frac{A_d}{A_s},
    \label{eq:supp_proportionality}
\end{equation}
where $A$ is nonuniversal.
This is the motivation for the comparison in Fig.~2\textbf{d} of the
main text.  Across different separations and the three system sizes
$L=6$, $8$, and $10$, the two independently constructed
symmetry-resolved correlations collapse approximately onto a common
proportional relation. 

\subsection{Additional data for spin and dimer correlation functions}

\begin{figure}[]
	\centering
	\includegraphics[width=1\linewidth]{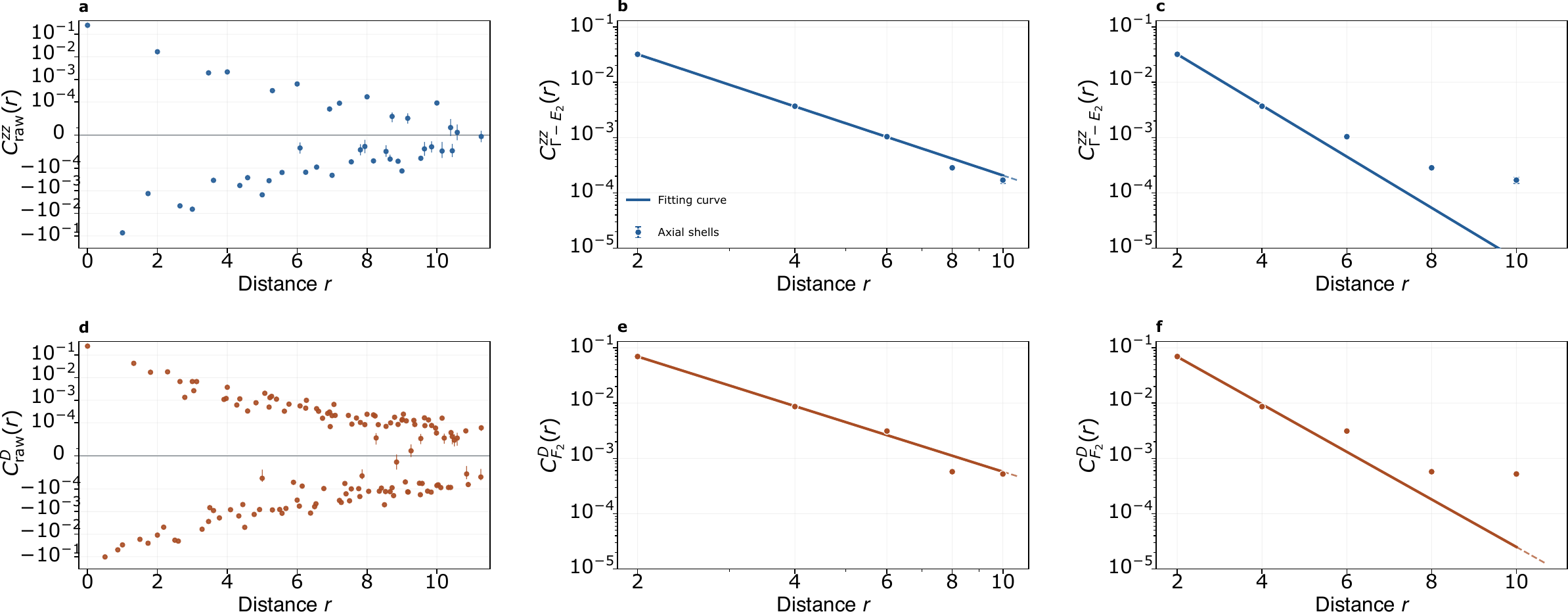}\\
	\caption{\textbf{Real-space spin and dimer correlations and their effective
scaling behavior for $L=10$.}
\textbf{a}, Symmetry-unresolved spin correlation
$C_{\rm raw}^{zz}(r)$, shown on a signed symmetric-logarithmic y-axis scale.  \textbf{b}, \textbf{c}, Symmetry-resolved $\Gamma$-$E_2$ spin correlation
$C_{\Gamma-E_2}^{zz}(r)$ fitted to a power-law form
$A r^{-p}$ and an exponential form $A'e^{-r/\xi}$, respectively.
The fits use the axial separations $r=2,4,6,8,10$ and yield
$p=3.13$ with $R^2=0.99997$ for the power-law fit and
$\xi=0.94$ with $R^2=0.99943$ for the exponential fit.
\textbf{d}, Symmetry-unresolved connected dimer correlation
$C_{\rm raw}^{D}(r)$, also shown on a signed symmetric-logarithmic
scale.
\textbf{e}, \textbf{f},  Symmetry-resolved $M$-$F_2$ dimer correlation
$C_{F_2}^{D}(r)$ fitted to the same power-law and exponential forms,
giving $p=2.98$ with $R^2=0.99984$ and
$\xi=1.01$ with $R^2=0.99862$, respectively. The raw spin and dimer correlations use physical site and bond-midpoint
separations, respectively, whereas the symmetry-resolved correlations
use Bravais-cell displacements. }\label{figs4}
\end{figure}

For comparison with the symmetry-resolved channels, here we define
the raw real-space spin correlation by averaging over all physical
site pairs belonging to the same periodic-boundary-condition distance
shell,
\begin{equation}
    C_{\rm raw}^{zz}(r)
    =
    \frac{1}{N_r^{(s)}}
    \sum_{\substack{i,j\\
    d_{\rm PBC}(i,j)=r}}
    \langle S_i^z S_j^z\rangle ,
    \label{eq:supp_Craw_spin}
\end{equation}
where $N_r^{(s)}$ denotes the number of site pairs in the shell at
distance $r$. We also define the raw connected dimer correlation by averaging
over all physical nearest-neighbour bond pairs with the same
periodic-boundary-condition distance between their bond midpoints,
\begin{equation}
    C_{\rm raw}^{D}(r)
    =
    \frac{1}{N_r^{(b)}}
    \sum_{\substack{b,b'\\
    d_{\rm PBC}(\mathbf r_b,\mathbf r_{b'})=r}}
    \left[
        \langle B_b B_{b'}\rangle
        -
        \langle B_b\rangle
        \langle B_{b'}\rangle
    \right],
    \label{eq:supp_Craw_dimer}
\end{equation}
where $\mathbf r_b$ denotes the midpoint of bond $b$, and
$N_r^{(b)}$ is the number of bond pairs in the corresponding distance
shell.  

The raw spin and dimer correlations are shown in
Figs.~\ref{figs4}\textbf{a} and \textbf{d}, respectively. Both quantities exhibit pronounced oscillations between positive and
negative values as a function of distance $r$, together with an overall
decrease in magnitude toward longer distances. Such behavior reflects the superposition of different internal
sublattice or bond-symmetry components in the symmetry-unresolved
correlators, and the raw correlations therefore do not exhibit a clear scaling form. 

To further examine the scaling behavior of the correlation functions, we now consider the symmetry-resolved correlation functions $C^{zz}_{\Gamma-E_{2}} (\mathbf{r})$ and $C^{D}_{F_{2}} (\mathbf{r})$ defined by Eq.~(\ref{eq:supp_CGammaE2}) and~(\ref{eq:supp_CF2}), respectively. 
The results of $C^{zz}_{\Gamma-E_{2}} (\mathbf{r})$ for $L=10$ are shown in Figs.~\ref{figs4}\textbf{b} and \textbf{c}, with the power-law, i.e., Eq.~(\ref{eq:supp_common_scaling}), and expotential fits, respectively. We emphasize that for the fitting analysis, we restrict ourselves to the axial $r=2,4,6,8$ and $10$, which correspond to displacements by one to five Bravais unit cells
along the principal lattice directions.  Since the primitive Bravais
vectors have length $2$ in units of the nearest-neighbor spacing,
these are precisely the nonzero axial separations available on the
$L=10$ torus up to the half-system distance. Restricting the fits to these axial shells allows us to compare the
distance dependence along a fixed crystallographic direction and
avoids mixing inequivalent lattice directions, which can exhibit
different finite-size and short-distance corrections. 

As shown in Figs.~\ref{figs4}\textbf{b} and \textbf{c}, the power-law form provides the better fit. Similarly, based on the results of $C^{D}_{F_{2}} (\mathbf{r})$ shown in Figs.~\ref{figs4}\textbf{e} and \textbf{f}, the power-law form describes the dimer correlation function better than the exponential form within the same fitting window. For the power-law fits, the decay rates of $C^{zz}_{\Gamma-E_{2}} (\mathbf{r})$ and $C^{D}_{F_{2}} (\mathbf{r})$ are $2\Delta_{\Phi} = 3.13$ and $2.98$, respectively, which are close to one another.





\section{Gutzwiller-projected spin-liquid reference states and wave-function fidelity}
\label{sec:fidelity}

In this section, we provide the explicit construction of the three
Gutzwiller-projected fermionic reference states used in Fig.~3 of the
main text and describe the evaluation of their fidelities with the
optimized KagomeResNet wave functions. Following the notation of the
main text, the three reference states are labeled by
\begin{equation}
    \mu=\mathrm{Dirac},\qquad
    \mu=Z_2,\qquad
    \mu=Z_2^\ast .
    \label{eq:supp_reference_labels}
\end{equation}
Here, $\mu=\mathrm{Dirac}$ denotes the optimized extended $U(1)$
Dirac state of Ref.~\cite{PhysRevB.84.020407};
$\mu=Z_2$ denotes a finite-pairing $Z_2[0,\pi]\beta$ state obtained
by perturbing the same Dirac state as considered in
Ref.~\cite{PhysRevB.84.020407}; and $\mu=Z_2^\ast$ denotes the
fully optimized $Z_2[0,\pi]\beta$ resonating-valence-bond (RVB)
state of Ref.~\cite{PhysRevB.102.195106}.
The auxiliary-spinon boundary condition is denoted by $\nu$, so that
the corresponding projected reference state is written as
$|\Phi_{\mu,\nu}\rangle$.

\subsection{Unified fermionic parent Hamiltonian}

We use the fermionic parton representation
\begin{equation}
    \mathbf S_i
    =
    \frac{1}{2}
    f_{i\alpha}^{\dagger}
    \boldsymbol{\sigma}_{\alpha\beta}
    f_{i\beta},
\end{equation}
subject to the local single-occupancy constraint
\begin{equation}
    n_{i\uparrow}+n_{i\downarrow}=1.
\end{equation}
All three reference states in Eq.~(\ref{eq:supp_reference_labels})
can be generated from the same general real singlet-BCS parent
Hamiltonian,
\begin{equation}
\begin{split}
    H_{\mathrm{par}}
    =&
    \sum_{ij,\sigma}
    t_{ij}
    f_{i\sigma}^{\dagger}f_{j\sigma}
    \\
    &+
    \frac{1}{2}
    \sum_{ij}
    \left[
        \Delta_{ij}
        \left(
            f_{i\uparrow}^{\dagger}f_{j\downarrow}^{\dagger}
            -
            f_{i\downarrow}^{\dagger}f_{j\uparrow}^{\dagger}
        \right)
        +\mathrm{H.c.}
    \right],
    \label{eq:supp_parent}
\end{split}
\end{equation}
where $t_{ij}=t_{ji}$ and $\Delta_{ij}=\Delta_{ji}$ are real in the
gauges considered below. The case $i=j$ is included, allowing for
on-site particle-hole and pairing terms.

Introducing the Nambu spinor
\begin{equation}
    \psi_i
    =
    \begin{pmatrix}
        f_{i\uparrow}\\
        f_{i\downarrow}^{\dagger}
    \end{pmatrix},
\end{equation}
the parent Hamiltonian may equivalently be expressed in terms of
matrices
\begin{equation}
    U_{ij}
    =
    t_{ij}\tau_3+\Delta_{ij}\tau_1
    \label{eq:supp_U_general}
\end{equation}
in the $\tau_1$--$\tau_3$ plane of the SU(2) gauge space.

For a given reference family $\mu$ and auxiliary-spinon boundary
condition $\nu$, we denote the mean-field ground state of
Eq.~(\ref{eq:supp_parent}) by
$|\Psi_{\mathrm{MF}}^{(\mu,\nu)}\rangle$. The physical reference
state entering the fidelity calculation is
\begin{equation}
    |\Phi_{\mu,\nu}\rangle
    =
    P_G
    |\Psi_{\mathrm{MF}}^{(\mu,\nu)}\rangle,
    \label{eq:supp_projected_reference}
\end{equation}
where
\begin{equation}
    P_G
    =
    \prod_i
    \delta_{n_{i\uparrow}+n_{i\downarrow},1}
\end{equation}
projects onto exactly one spinon per kagome site.

Equation~(\ref{eq:supp_parent}) provides a common parent-Hamiltonian
framework for all three reference states. The specific hopping and
pairing amplitudes, however, must be implemented in the gauge
convention of the corresponding literature Ansatz. In particular,
the parametrization of Ref.~\cite{PhysRevB.102.195106} should not be
obtained by directly identifying its parameters with those of
Ref.~\cite{PhysRevB.84.020407}; the two Ans\"atze use different bond
sign conventions and the former additionally contains third-neighbour
RVB amplitudes.

\subsection{Dirac reference state: $\mu=\mathrm{Dirac}$}

For $\mu=\mathrm{Dirac}$, we use the optimized extended $U(1)$
Dirac state of Ref.~\cite{PhysRevB.84.020407}. In the gauge used
therein, the NN and second-neighbour mean-field matrices are
\begin{align}
    U_{ij}^{(1)}
    &=
    s_{ij}\chi_1\tau_3,
    \nonumber\\
    U_{ij}^{(2)}
    &=
    \nu_{ij}^{\mathrm I}\chi_2\tau_3,
    \label{eq:supp_dirac_ansatz}
\end{align}
where $s_{ij}=\pm1$ and $\nu_{ij}^{\mathrm I}=\pm1$ follow the
bond-sign convention of Fig.~1 of
Ref.~\cite{PhysRevB.84.020407}. The Ansatz has zero gauge flux
through the elementary triangles and $\pi$ flux through the
hexagons.

The NN hopping defines the overall energy scale,
\begin{equation}
    \chi_1=1.
\end{equation}
The optimized extended Dirac state reported in
Ref.~\cite{PhysRevB.84.020407} has
\begin{equation}
    \chi_2=-0.0186(2),
    \qquad
    \mu_{\mathrm I}=-0.5124(5),
    \qquad
    \Delta_2=\zeta=0,
    \label{eq:supp_dirac_parameters}
\end{equation}
where $\mu_{\mathrm I}$ denotes the uniform on-site chemical
potential of the mean-field Hamiltonian. We use the subscript
``I'' here to distinguish this parameter from the reference-state
label $\mu$ employed throughout the main text.

Because the state is number conserving and the projected wave
function is constructed at fixed half filling, the uniform
chemical-potential term does not change the occupied
single-particle orbitals. The projected reference state is therefore
fully determined by the hopping Ansatz in
Eq.~(\ref{eq:supp_dirac_ansatz}) at half filling.

Ref.~\cite{PhysRevB.84.020407} performed the variational calculation
on a $432$-site cluster with mixed periodic-antiperiodic spinon
boundary conditions and obtained
\begin{equation}
    \frac{E_{\mathrm{Dirac}}}{N}
    =
    -0.42872(1)J.
    \label{eq:supp_dirac_energy}
\end{equation}

\subsection{Finite-pairing $Z_2$ reference state: $\mu=Z_2$}

The reference state denoted by $\mu=Z_2$ in the main text is a
finite-pairing $Z_2[0,\pi]\beta$ state in the immediate variational
neighbourhood of the Dirac state of
Ref.~\cite{PhysRevB.84.020407}. Its mean-field matrices are
\begin{align}
    U_{ii}^{(0)}
    &=
    \mu_{\mathrm I}\tau_3+\zeta\tau_1,
    \nonumber\\
    U_{ij}^{(1)}
    &=
    s_{ij}\chi_1\tau_3,
    \nonumber\\
    U_{ij}^{(2)}
    &=
    \nu_{ij}^{\mathrm I}
    \left(
        \chi_2\tau_3+\Delta_2\tau_1
    \right),
    \nonumber\\
    U_{ij}^{(3)}
    &=0.
    \label{eq:supp_iqbal_z2_ansatz}
\end{align}
A nonzero $\Delta_2$ generates noncommuting SU(2) fluxes and breaks
the emergent $U(1)$ gauge structure of the Dirac Ansatz down to
$Z_2$.

The specific finite-pairing reference used in Fig.~3 is defined by
\begin{equation}
\begin{split}
    \chi_1 &=1,\\
    \chi_2 &=-0.0186,\\
    \mu_{\mathrm I} &=-0.5124,\\
    \Delta_2 &=0.05,\\
    \zeta &=0.1780.
\end{split}
\label{eq:supp_iqbal_z2_parameters}
\end{equation}
For $\Delta_2=0.05$, Ref.~\cite{PhysRevB.84.020407} reports the
corresponding optimized on-site pairing
$\zeta=0.1780(2)$. Within the statistical resolution of the VMC
calculation, introducing this finite pairing produces no detectable
change in the variational energy relative to the Dirac state.
Therefore, on the same $N=432$ cluster,
\begin{equation}
    \frac{E_{Z_2}}{N}
    \simeq
    -0.42872(1)J,
    \label{eq:supp_iqbal_z2_energy}
\end{equation}
where Ref.~\cite{PhysRevB.84.020407} does not quote a separate
independent energy value for this particular finite-pairing point.

We stress that the state defined by
Eq.~(\ref{eq:supp_iqbal_z2_parameters}) is not the fully optimized
minimum of the $Z_2[0,\pi]\beta$ variational family. Upon complete
stochastic-reconfiguration optimization,
Ref.~\cite{PhysRevB.84.020407} finds
$\Delta_2,\zeta\rightarrow0$, such that the optimized state returns
to the $U(1)$ Dirac state. We retain the finite-pairing state as an
independent reference in order to test whether the optimized NQS
shows an enhanced affinity to an explicit
$U(1)\rightarrow Z_2$ pairing perturbation whose energy is nearly
indistinguishable from that of the Dirac state.

\subsection{Optimized $Z_2$ reference state: $\mu=Z_2^\ast$}

For the second $Z_2$ reference, denoted by $\mu=Z_2^\ast$, we use
the fully optimized $Z_2[0,\pi]\beta$ RVB state of
Ref.~\cite{PhysRevB.102.195106}. This Ansatz is defined in a
different gauge from that of
Ref.~\cite{PhysRevB.84.020407} and contains RVB amplitudes up to
third-neighbour bonds.

Following Ref.~\cite{PhysRevB.102.195106}, we introduce
\begin{equation}
    \mathbf n_\phi
    =
    (\sin\phi,0,\cos\phi),
\end{equation}
such that
\begin{equation}
    \mathbf n_\phi\cdot\boldsymbol{\tau}
    =
    \sin\phi\,\tau_1+\cos\phi\,\tau_3.
\end{equation}
With the NN amplitude fixed to unity, the parent matrices are
\begin{equation}
    U_{ij}
    =
    \begin{cases}
    -\mu_{\mathrm Z}\,
    \mathbf n_{\phi_1}\cdot\boldsymbol{\tau},
    & i=j,\\[1mm]
    -s_{ij}^{\mathrm Z}\tau_3,
    & \langle ij\rangle_1,\\[1mm]
    -s_{ij}^{\mathrm Z}\rho\,
    \mathbf n_{\phi_2}\cdot\boldsymbol{\tau},
    & \langle ij\rangle_2,\\[1mm]
    -s_{ij}^{\mathrm Z}\eta\,
    \mathbf n_{\phi_3}\cdot\boldsymbol{\tau},
    & \langle ij\rangle_3.
    \end{cases}
    \label{eq:supp_zhang_ansatz}
\end{equation}
Here $s_{ij}^{\mathrm Z}=\pm1$ follows the six-site spinon-unit-cell
gauge shown in Fig.~1 of
Ref.~\cite{PhysRevB.102.195106}. The spinon unit cell is doubled
along $\mathbf a_1$ and therefore contains six kagome sites.

In terms of the hopping and pairing amplitudes of the unified parent
Hamiltonian in Eq.~(\ref{eq:supp_parent}), the second- and
third-neighbour terms are
\begin{align}
    t_{ii}
    &=
    -\mu_{\mathrm Z}\cos\phi_1,
    &
    \Delta_{ii}
    &=
    -\mu_{\mathrm Z}\sin\phi_1,
    \nonumber\\
    t_{ij}^{(2)}
    &=
    -s_{ij}^{\mathrm Z}\rho\cos\phi_2,
    &
    \Delta_{ij}^{(2)}
    &=
    -s_{ij}^{\mathrm Z}\rho\sin\phi_2,
    \nonumber\\
    t_{ij}^{(3)}
    &=
    -s_{ij}^{\mathrm Z}\eta\cos\phi_3,
    &
    \Delta_{ij}^{(3)}
    &=
    -s_{ij}^{\mathrm Z}\eta\sin\phi_3.
    \label{eq:supp_zhang_cartesian}
\end{align}
Thus, $\eta$ is the magnitude of the third-neighbour RVB parameter
and should not be identified with the on-site pairing $\zeta$ in
Eq.~(\ref{eq:supp_iqbal_z2_ansatz}). Similarly, $\rho$ cannot be
identified directly with the Iqbal parameter $\chi_2$ without an
explicit gauge transformation of the complete bond-sign structure.

For the three system sizes studied in the present work, we use the
finite-size optimized parameters shown in Fig.~3 of
Ref.~\cite{PhysRevB.102.195106}. The values used in our numerical
implementation are
\begin{table}[t]
\caption{Parameters of the $Z_2^\ast$ reference state,
corresponding to the finite-size optimized $Z_2[0,\pi]\beta$
Ansatz of Ref.~\cite{PhysRevB.102.195106}.}
\label{tab:supp_zhang_parameters}
\begin{ruledtabular}
\begin{tabular}{c c c c c c c}
$L$ &
$\mu_{\mathrm Z}$ &
$\rho$ &
$\eta$ &
$\phi_1$ &
$\phi_2$ &
$\phi_3$\\
\hline
6  & 0.4884 & 0.3126 & 0.3428 & 2.0412 & 2.8204 & 2.8421\\
8  & 0.5534 & 0.3577 & 0.3920 & 2.5210 & 2.9475 & 2.9692\\
10 & 0.5739 & 0.3788 & 0.4137 & 2.6851 & 3.0009 & 3.0246
\end{tabular}
\end{ruledtabular}
\end{table}
The optimized parameters are presented graphically rather than
tabulated numerically in Ref.~\cite{PhysRevB.102.195106};
the values in Table~\ref{tab:supp_zhang_parameters} were therefore
obtained by digitizing the published Fig.~3. The retained digits are
used only to define the numerical reference states reproducibly and
should not be interpreted as table-level precision of the original
publication.

The corresponding variational energies reported directly in
Table~I of Ref.~\cite{PhysRevB.102.195106} are
\begin{table}[t]
\caption{Variational energy density of the optimized
$Z_2^\ast$ reference state reported in
Ref.~\cite{PhysRevB.102.195106}.}
\label{tab:supp_zhang_energies}
\begin{ruledtabular}
\begin{tabular}{c c c}
$L$ & $N=3L^2$ & $E_{Z_2^\ast}/N$\\
\hline
6  & 108 & $-0.428899(1)J$\\
8  & 192 & $-0.428783(1)J$\\
10 & 300 & $-0.428753(1)J$\\
18 & 972 & $-0.428737(1)J$
\end{tabular}
\end{ruledtabular}
\end{table}
The $L=18$ cluster is the largest size for which
Ref.~\cite{PhysRevB.102.195106} reports well-converged optimized
RVB parameters. At this size, the corresponding optimized $U(1)$
state has energy density $-0.428719(1)J$, while the optimized
$Z_2$ state has $-0.428737(1)J$. Thus, within this projected-RVB
variational family, the $Z_2$ state lies slightly below the optimized
$U(1)$ state in energy.

The three reference states used in Fig.~3 are therefore
complementary. The state $\mu=\mathrm{Dirac}$ provides an optimized
extended $U(1)$ Dirac reference; $\mu=Z_2$ probes an explicit
finite-pairing perturbation in the immediate neighbourhood of this
Dirac state; and $\mu=Z_2^\ast$ provides an independently optimized
and energetically competitive $Z_2[0,\pi]\beta$ reference with
additional third-neighbour variational freedom.

\subsection{Auxiliary-spinon boundary conditions}

The physical kagome Heisenberg Hamiltonian and the KagomeResNet NQS
are always defined with periodic boundary conditions. The boundary
conditions discussed here apply only to the auxiliary fermions used
to construct the Gutzwiller-projected reference states.

For an $L\times L$ torus, we define
\begin{equation}
    f_{\mathbf r+L\mathbf a_1}
    =
    e^{i\theta_1}
    f_{\mathbf r},
    \qquad
    f_{\mathbf r+L\mathbf a_2}
    =
    e^{i\theta_2}
    f_{\mathbf r},
    \label{eq:supp_spinon_bc}
\end{equation}
with $\theta_{1,2}=0$ or $\pi$. We denote periodic and
antiperiodic boundary conditions by P and A, respectively.

Following the mixed periodic-antiperiodic boundary conditions used
in the original VMC calculations
Refs.~\cite{PhysRevB.84.020407,PhysRevB.102.195106}, the fidelity
comparison in the main text is restricted to
\begin{equation}
    \nu\in\{\mathrm{PA},\mathrm{AP}\}.
    \label{eq:supp_mixed_sectors}
\end{equation}
Here, the first and second letters refer to the spinon boundary
conditions along $\mathbf a_1$ and $\mathbf a_2$, respectively.

For $\mu=\mathrm{Dirac}$, the number-conserving mean-field state is
used only when the half-filled spinon spectrum forms a nondegenerate
closed shell. For the system sizes considered here, the two mixed
sectors satisfy
\begin{table}[t]
\caption{Closed-shell structure of the two mixed spinon boundary
conditions for the Dirac reference state.}
\label{tab:supp_closed_shell}
\begin{ruledtabular}
\begin{tabular}{c c c}
$L$ & PA & AP\\
\hline
6  & open shell   & closed shell\\
8  & closed shell & closed shell\\
10 & open shell   & closed shell
\end{tabular}
\end{ruledtabular}
\end{table}
An open-shell sector is excluded because the degenerate
single-particle level at the Fermi energy does not define a unique
half-filled Slater determinant without an additional prescription
for the zero-mode occupations.

For the paired references $\mu=Z_2$ and $\mu=Z_2^\ast$, the BdG
ground state is well defined in both mixed sectors, and both PA and
AP are retained. Consequently, the quantity reported in Fig.~3 of
the main text is
\begin{equation}
    F_\mu^{\max}
    =
    \max_{\nu\in{\cal B}_{\mu}(L)}
    F_{\mu,\nu},
    \label{eq:supp_Fmax}
\end{equation}
where ${\cal B}_{\mu}(L)$ denotes the admissible set of mixed
boundary sectors for reference state $\mu$. For
$\mu=\mathrm{Dirac}$, ${\cal B}_{\mu}(L)$ contains only
nondegenerate closed-shell sectors; for
$\mu=Z_2$ and $Z_2^\ast$, both mixed sectors are included.

\subsection{Projected amplitudes and Monte Carlo evaluation of the fidelity}

For the number-conserving reference $\mu=\mathrm{Dirac}$, the
mean-field ground state is obtained by filling the lowest $N/2$
single-particle orbitals for each spin species. For a spin
configuration $\mathbf s$ with sets of up- and down-spin sites
$\mathcal U$ and $\mathcal D$, respectively, the Gutzwiller-projected
amplitude can be expressed as the product of the corresponding
Slater determinants,
\begin{equation}
    \Phi_{\mathrm{Dirac},\nu}(\mathbf s)
    \propto
    (-1)^{P(\mathbf s)}
    \det M_{\mathcal U}^{(\nu)}
    \det M_{\mathcal D}^{(\nu)} ,
    \label{eq:supp_dsl_amplitude}
\end{equation}
where $(-1)^{P(\mathbf s)}$ accounts for the fermionic reordering
required to express the projected state in the spin-configuration
basis used for the NQS.

For $\mu=Z_2$ and $Z_2^\ast$, diagonalization of the corresponding
BdG Hamiltonian yields the BCS pair matrix
$F^{(\mu,\nu)}$. After projection to
$N_\uparrow=N_\downarrow=N/2$, the amplitude is
\begin{equation}
    \Phi_{\mu,\nu}(\mathbf s)
    \propto
    (-1)^{P(\mathbf s)}
    \det
    F_{\mathcal U,\mathcal D}^{(\mu,\nu)},
    \qquad
    \mu=Z_2,Z_2^\ast .
    \label{eq:supp_bcs_amplitude}
\end{equation}
Thus, all three projected references and the optimized NQS are
evaluated in exactly the same physical spin basis.

For each admissible pair $(\mu,\nu)$, the normalized squared
many-body fidelity is
\begin{equation}
    F_{\mu,\nu}
    =
    \frac{
    \left|
        \langle
        \Psi_{\theta}^{\mathrm{opt}}
        |
        \Phi_{\mu,\nu}
        \rangle
    \right|^2
    }{
    \langle
        \Psi_{\theta}^{\mathrm{opt}}
        |
        \Psi_{\theta}^{\mathrm{opt}}
    \rangle
    \,
    \langle
        \Phi_{\mu,\nu}
        |
        \Phi_{\mu,\nu}
    \rangle
    }.
    \label{eq:supp_fidelity}
\end{equation}
This is the same quantity defined in the main text.

We evaluate Eq.~(\ref{eq:supp_fidelity}) using configurations sampled
from the optimized NQS probability distribution
\begin{equation}
    p_{\theta}(\mathbf s)
    =
    \frac{
        |\Psi_{\theta}^{\mathrm{opt}}(\mathbf s)|^2
    }{
        \langle
        \Psi_{\theta}^{\mathrm{opt}}
        |
        \Psi_{\theta}^{\mathrm{opt}}
        \rangle
    }.
\end{equation}
For each reference state, we define the amplitude ratio
\begin{equation}
    R_{\mu,\nu}(\mathbf s)
    =
    \frac{
        \Phi_{\mu,\nu}(\mathbf s)
    }{
        \Psi_{\theta}^{\mathrm{opt}}(\mathbf s)
    }.
    \label{eq:supp_ratio}
\end{equation}
It then follows that
\begin{equation}
    \left\langle
        R_{\mu,\nu}
    \right\rangle_{p_\theta}
    =
    \frac{
        \langle
        \Psi_{\theta}^{\mathrm{opt}}
        |
        \Phi_{\mu,\nu}
        \rangle
    }{
        \langle
        \Psi_{\theta}^{\mathrm{opt}}
        |
        \Psi_{\theta}^{\mathrm{opt}}
        \rangle
    },
\end{equation}
and
\begin{equation}
    \left\langle
        |R_{\mu,\nu}|^2
    \right\rangle_{p_\theta}
    =
    \frac{
        \langle
        \Phi_{\mu,\nu}
        |
        \Phi_{\mu,\nu}
        \rangle
    }{
        \langle
        \Psi_{\theta}^{\mathrm{opt}}
        |
        \Psi_{\theta}^{\mathrm{opt}}
        \rangle
    }.
\end{equation}
Therefore,
\begin{equation}
    F_{\mu,\nu}
    =
    \frac{
        \left|
        \left\langle
            R_{\mu,\nu}
        \right\rangle_{p_\theta}
        \right|^2
    }{
        \left\langle
            |R_{\mu,\nu}|^2
        \right\rangle_{p_\theta}
    }.
    \label{eq:supp_fidelity_estimator}
\end{equation}
Neither the NQS nor the projected reference state therefore needs to
be normalized explicitly.

The fidelity provides information complementary to the
symmetry-resolved correlation functions discussed in the preceding
section. Whereas the latter probe selected long-distance operators,
$F_{\mu,\nu}$ compares the complete many-body amplitude and phase
structure of the optimized NQS with a given projected spin-liquid
reference state.

As shown in Fig.~3 of the main text, the finite-pairing state
$\mu=Z_2$ has a substantially smaller maximum fidelity than
$\mu=\mathrm{Dirac}$, despite the two reference states having
indistinguishable variational energies at the resolution reported in
Ref.~\cite{PhysRevB.84.020407}. The more competitive reference
$\mu=Z_2^\ast$ has a fidelity close to that of the Dirac state for
$L=6$, but its $F_{Z_2^\ast}^{\max}$ becomes systematically smaller
than $F_{\mathrm{Dirac}}^{\max}$ with increasing system size.
Importantly, the $Z_2^\ast$ reference was found in
Ref.~\cite{PhysRevB.102.195106} to have a slightly lower
projected-VMC energy than the corresponding optimized $U(1)$ state.
The fidelity comparison is therefore not simply tracking the
variational-energy ordering within the projected-fermion family, but
provides an independent wave-function-level diagnostic supporting
the Dirac interpretation of the optimized KagomeResNet ground state.

\bibliography{reference_ka}